\documentclass[english,aps,a4paper,prd,groupedaddress,preprintnumbers,nofootinbib,preprint,floatfix,showpacs]{revtex4-2}

\usepackage{hyperref}
\usepackage{amsmath,bbm,latexsym,amssymb}
\usepackage{graphicx}
\usepackage{rotating}
\usepackage{color} 
\usepackage{multirow} 
\usepackage[T1]{fontenc} 
\usepackage{lmodern}
\makeatletter

\usepackage{booktabs}
\usepackage{mathrsfs}   
\usepackage{slashed}     
\usepackage{url}
\usepackage{units}
\usepackage{wasysym }
\usepackage{comment}

\usepackage{float}
\newcommand{\bea}{\begin{eqnarray}}
\newcommand{\eea}{\end{eqnarray}}

\newcommand{\be}{\begin{equation}}
\newcommand{\ee}{\end{equation}}
\def\beq{\begin{equation}}
\def\eeq{\end{equation}}
\newcommand{\ba}{\begin{eqnarray}}
\newcommand{\ea}{\end{eqnarray}}

 \def\ifmath#1{\relax\ifmmode #1\else $#1$\fi}
\def\eq#1{Eq.~(\ref{#1})}
\def\eqs#1#2{Eqs.~(\ref{#1}) and (\ref{#2})}

\def\sgi  {s_{\gamma_1}}
\def\cgi  {c_{\gamma_1}}
\def\sgii  {s_{\gamma_2}}
\def\cgii  {c_{\gamma_2}}
\def\sai  {s_{\alpha_1}}
\def\cai  {c_{\alpha_1}}
\def\saii  {s_{\alpha_2}}
\def\caii  {c_{\alpha_2}}
\def\saiii  {s_{\alpha_3}}
\def\caiii  {c_{\alpha_3}}

\makeatother

\usepackage{babel}

\newcommand{\Z}[1]{\ensuremath{\mathbbm{Z}_{#1}}} 

\makeatletter
\renewcommand*\env@matrix[1][\arraystretch]{%
  \edef\arraystretch{#1}%
  \hskip -\arraycolsep
  \let\@ifnextchar\new@ifnextchar
  \array{*\c@MaxMatrixCols c}}
\makeatother

\newcommand{\AddrCFTP}{%
 Departamento de F\'\i sica and CFTP, Instituto Superior T\'ecnico\\
 Universidade de Lisboa, 
          Av. Rovisco Pais 1, 1049-001 Lisboa, Portugal }

\def\gsim{\raise0.3ex\hbox{$\;>$\kern-0.75em\raise-1.1ex\hbox{$\sim\;$}}}
\def\lsim{\raise0.3ex\hbox{$\;<$\kern-0.75em\raise-1.1ex\hbox{$\sim\;$}}}

\allowdisplaybreaks

\begin{document}

\preprint{CFTP/22-004}

\title{BFB conditions on a class of symmetry constrained 3HDM} 
\author{Rafael Boto}%
\email{rafael.boto@tecnico.ulisboa.pt}
\affiliation{\AddrCFTP}
\author{Jorge C.~Rom\~{a}o}\email{jorge.romao@tecnico.ulisboa.pt}
\affiliation{\AddrCFTP} 
 \author{Jo\~{a}o  P.~Silva} \email{jpsilva@cftp.ist.utl.pt}
 \affiliation{\AddrCFTP}

\today

\begin{abstract}
We study the bounded from below (BFB) conditions on  a class of three
Higgs doublet models (3HDM) constrained by the symmetry groups $U(1)\times U(1)$,
$U(1)\times \Z2$, and $\Z2\times \Z2$.
These constraints must be implemented on both the neutral (BFB-n)
and charged (BFB-c) directions.
The exact necessary and sufficient BFB conditions are unknown in the
$\Z2\times \Z2$ case.
We develop a general strategy using lower bounds to find  sufficient conditions
for BFB-n and BFB-c and apply it to these symmetries.
In addition, we investigate the concern that the use of safe sufficient conditions can
ignore valid points which would yield distinct physical consequences.
This is done by performing a full phenomenological simulation of
the $U(1)\times U(1)$ and $U(1)\times \Z2$ models,
where exact necessary and sufficient BFB conditions are possible.
We look specifically at the points allowed by exact solutions but precluded by safe lower bounds.
We found no evidence of remarkable new effects, 
partly reassuring the use of the lower bounds we propose here,
for those potentials where no exact necessary and sufficient BFB conditions are known.
\end{abstract}

\maketitle

\newpage 

\section{\label{sec:intro}Introduction}

It is widely accepted that there must be Physics Beyond the Standard Model (SM).
One reason concerns the necessity to provide new CP-violating phases and
a stronger phase transition in order to drive baryogenesis.
A second reason concerns the necessity to find one or more new particles that
describe Dark Matter.
There are also the issues of explaining neutrino masses and the possibility that
these might have a Majorana character, or of providing an explanation
for the observed mass hierarchies and mixing matrices.
 
The large majority of models addressing these issues include extended scalar sectors.
Nevertheless, many times,
perhaps because it is a very difficult problem, the issue of having
a potential bounded from below (BFB) or guaranteeing that the vacuum is indeed
a global (not just local) minimum is ignored.
Occasionally,
some BFB conditions are included without stressing whether such conditions are
necessary, sufficient or both.
And, most articles addressing this problem concentrate on BFB conditions analyzing
only vacua along the neutral directions; that is, vacua which do not
break electric charge.

However,
Ivanov and Faro \cite{Faro:2019vcd} showed,
using the specific case of a $U(1) \times U(1)$ symmetric
three Higgs doublet model (3HDM),
that one can have a minimum of the potential which satisfies the condition for
bounded for below along charge preserving directions,
but is still unbounded from below along the charge breaking (CB) directions.
They then proceeded to establish necessary and sufficient conditions for
BFB along both neutral (BFB-n) and charge breaking (BFB-c) directions,
for the specific case of the $U(1) \times U(1)$ 3HDM.
Faro \cite{Faro:2019} extended this analysis to the $U(1) \times \Z2$ symmetric 3HDM;
an extension which is unpublished and little known.
For example,
the recent \cite{Das:2021oik},
which has this potential, does not use these complete necessary and sufficient
BFB conditions. We reproduce this result here.

Surprisingly,
there are no known necessary and sufficient conditions for BFB
for such a simple and classical model as the $\Z2 \times \Z2$ 3HDM.
This model was first proposed by Weinberg in \cite{Weinberg:1976hu},
in order to have CP violation in the scalar sector,
without exhibiting flavour changing neutral scalar couplings.
The best result has been derived in \cite{Grzadkowski:2009bt},
which has the necessary and sufficient conditions for BFB-n and
sufficient conditions for BFB-c.

A first aim of this article is to
present a method to derive BFB-n and BFB-c sufficient conditions,
in cases where necessary \textit{and} sufficient conditions are not
available through other techniques.
The method hinges on finding a potential which lies lower than the potential
desired, and for which one can apply the copositivity conditions
of Klimenko \cite{Klimenko:1984qx} and Kannike \cite{Kannike:2012pe}
in order to find BFB conditions for that new potential.
We apply this method to the 
3HDM with the symmetries $U(1)\times U(1)$,
$U(1)\times \Z2$, and $\Z2\times \Z2$.
The method can be applied to a generality of other cases;
see for example \cite{Carrolo:2022oyg}.

If one uses necessary (but not sufficient) conditions for BFB, one is basing
the analysis on some potentials which are unphysical. Conversely, if one
uses sufficient (but not necessary) conditions for BFB, one is excluding
perfectly good potentials, running the risk that these have some special
features, potentially ignoring interesting new physics signals.
Although we are unaware on any specific case in which this has happened,
there is an ever worse possibility; that the potentials which pass sufficient BFB
conditions are all excluded, while those which are physical but do not pass
such sufficient conditions are still allowed. One would thus erroneously
consider as excluded a perfectly viable model.

It is interesting to address the latter concern,
given that we have both the sufficient BFB conditions (BFB-n \textit{and} BFB-c)
for the $U(1)\times U(1)$ and $U(1)\times \Z2$ 3HDM using our method,
and also the complete necessary and sufficient BFB conditions 
for these cases.
We can thus see if the points which pass the necessary and sufficient
conditions but do not pass the more stringent sufficient conditions
hold some special physically observable property.
This is the second aim of this article.

Our article is organized as follows.
We define the notation for the scalar potential in Section~\ref{sec:scalar_pot}.
For the cases of $U(1)\times U(1)$ and $U(1)\times \Z2$,
we show in Sections~\ref{sec:BFB-U1xU1} and \ref{sec:BFB-U1xZ2}, respectively,
the necessary and sufficient conditions for BFB,
the adaptation to these cases of the sufficient
conditions in \cite{Grzadkowski:2009bt},
and the sufficient conditions derived with our method.
The $\Z2 \times \Z2$ case is discussed in Section~\ref{sec:BFB-Z2xZ2}.

In Section~\ref{sec:scalar} we introduce the rotation into the scalar mass bases,
thus allowing a parameterization of the potential parameters in terms of
physical quantities.
This is presented for the three cases in complete form in Appendix~\ref{app:lambdas}.
Next, we consider in Section~\ref{sec:Yukawas}  complemented by
Appendix~\ref{app:Types} the Yukawa sector,
showing the symmetry and parameterization of the five types of
models which preclude flavour changing neutral scalar exchanges,
introducing the so-called $k$-notation in Section~\ref{sec:kappas}.
The scan strategy and results are discussed in Sections~\ref{ref:scan}
and \ref{sec:results}, respectively.
We present our conclusions in Section~\ref{sec:conclusions}.
We relegate two other technical details to Appendix~\ref{app:STU} and \ref{app:Unitarity}.

\section{\label{sec:scalar_pot}The Potential}  

We consider the potential defined by,
\begin{equation}\label{Z3potential}
    V=V_{2}+V_{4}\, .
\end{equation}
As for now we are only interested in the BFB conditions, we just consider
the quartic terms invariant under the relevant group $G$.
All symmetry constrained three Higgs doublet model (3HDM) potentials
have a piece invariant under rephasings; that is, invariant under
$U(1) \times U(1)$:~\footnote{Invariance under hypercharge guarantees that requiring
invariance under rephasings of two scalar fields implies automatically invariance under
rephasing of the third field.}
\begin{align}
V_{4, \textrm{RI}}=&
\lambda_1(\phi_1^\dagger\phi_1)^2
+\lambda_2(\phi_2^\dagger\phi_2)^2
+\lambda_3(\phi_3^\dagger\phi_3)^2+
\lambda_4(\phi_1^\dagger\phi_1)(\phi_2^\dagger\phi_2)
+\lambda_5(\phi_1^\dagger\phi_1)(\phi_3^\dagger\phi_3)\nonumber\\[8pt]
& 
+\lambda_6(\phi_2^\dagger\phi_2)(\phi_3^\dagger\phi_3)
+\lambda_7(\phi_1^\dagger\phi_2)(\phi_2^\dagger\phi_1)
+\lambda_8(\phi_1^\dagger\phi_3)(\phi_3^\dagger\phi_1)
+\lambda_9(\phi_2^\dagger\phi_3)(\phi_3^\dagger\phi_2)\, .
\label{U1U1quartic}
\end{align}
The rephasing invariant quartic couplings can be written alternatively as
\begin{equation}
V_{4, \textrm{RI}}=
V_N + V_{CB}\, ,
\label{U1U1quartic_2}
\end{equation}
where
\begin{align}
  V_N=&\frac{\lambda_{11}}{2}(\phi_1^\dagger\phi_1)^2 +
\frac{\lambda_{22}}{2}(\phi_2^\dagger\phi_2)^2+
\frac{\lambda_{33}}{2}(\phi_3^\dagger\phi_3)^2+
\lambda_{12}(\phi_1^\dagger\phi_1) (\phi_2^\dagger\phi_2)+
\lambda_{13}(\phi_1^\dagger\phi_1)
(\phi_3^\dagger\phi_3)
\nonumber\\[4pt]
&+
\lambda_{23}(\phi_2^\dagger\phi_2) (\phi_3^\dagger\phi_3)\, ,
\label{eq:VN}\\[4pt]
V_{CB}=& \lambda'_{12} z_{12}+\lambda'_{13} z_{13}+
\lambda'_{23} z_{23}\, ,
\label{eq:VCB}
\end{align}
and \cite{Faro:2019vcd}
\begin{equation}
  \label{eq:25}
  z_{ij}= (\phi_i^\dagger\phi_i) (\phi_j^\dagger\phi_j)
  - (\phi_i^\dagger\phi_j) (\phi_j^\dagger\phi_i) \quad \text{(no sum)}\, .
\end{equation}
Notice that we have always
\begin{equation}
  \label{eq:6}
  0\le z_{ij} \le r_i r_j\, ,
\end{equation}
where $r_k = |\phi_k|^2\ (k=1,2,3)$ -- see also Eq.~\eqref{eq:22} below.
With these conventions the relation between the two notations is
\begin{align}
  \label{eq:1}
 & \lambda_{11}\to 2 \lambda_1,\
  \lambda_{22}\to 2 \lambda_2,\
  \lambda_{33}\to 2 \lambda_3,\
  \lambda_{12}\to \lambda_4+\lambda_7,\
  \lambda_{13}\to \lambda_5+\lambda_8,\\[+2mm]
 & \lambda_{23}\to \lambda_6+\lambda_9,\ 
  \lambda'_{12}\to -\lambda_7,\
  \lambda'_{13}\to -\lambda_8,\
  \lambda'_{23}\to -\lambda_9 .
\end{align}

Given a potential invariant under a group $G$,
its quartic part may be written as
\begin{equation}
V_4 = V_{4, \textrm{RI}} + V_G\, ,
\end{equation}
where $V_{4, \textrm{RI}}$ is the rephasing invariant piece
of Eqs.~\eqref{U1U1quartic}-\eqref{U1U1quartic_2}, common to all potentials,
while $V_G$ is the rephasing non-invariant part of the quartic potential
that depends on the
group. The groups $G = U(1)\times U(1)$ (for which, obviously $V_G=0$),
$U(1)\times \Z2$, and $\Z2\times \Z2$,
are discussed in detail in the corresponding sections below.

\section{BFB conditions in the $U(1)\times U(1)$ case}
\label{sec:BFB-U1xU1}

Let us consider the $U(1)\times U'(1)$ transformation\footnote{When convenient to distinguish
the two $U(1)$'s, we will denote the second one by a prime.}
\begin{align}
U(1) &:   &   \phi_1 &\rightarrow e^{i \theta}\phi_1
&   \phi_2 &\rightarrow \phi_2
&   \phi_3 &\rightarrow \phi_3\, ,
\label{U(1)_scalar}
\\
U'(1) &:   &   \phi_1 &\rightarrow \phi_1
&   \phi_2 &\rightarrow e^{i \theta'}\phi_2
&   \phi_3 &\rightarrow \phi_3\, .
\label{U'(1)_scalar}
\end{align}
where the transformations are to be implemented for all $\theta$ and $\theta'$.
This is the simplest case because the symmetry forces $V_G=0$, so
\begin{equation}
  \label{eq:15a}
  V_4=V_N + V_{CB}\, ,
\end{equation}
given in Eqs.~(\ref{eq:VN}) and (\ref{eq:VCB}).

\subsection{Necessary and sufficient conditions for BFB}
\label{subsec:U1U1_BFB_NS}

The necessary and
sufficient conditions for BFB of the potential
for this case were found by Faro and
Ivanov \cite{Faro:2019vcd}. They can be enunciated in three steps.
For these,
we use gauge invariance to parameterize the (vevs of the) doublets
as \cite{Faro:2019vcd},
\begin{equation}
  \label{eq:22}
  \phi_1=\sqrt{r_1}
  \begin{pmatrix}
    0\\
    1
  \end{pmatrix},\quad
  \phi_2=\sqrt{r_2}
  \begin{pmatrix}
    \sin(\alpha_2)\\
    \cos(\alpha_2) e^{i \beta_2}
  \end{pmatrix},\quad
  \phi_3=\sqrt{r_3} e^{i \gamma}
  \begin{pmatrix}
    \sin(\alpha_3)\\
    \cos(\alpha_3) e^{i \beta_3}
  \end{pmatrix}  .
\end{equation}
%


\subsubsection{Step 1}

The potential along the neutral directions, $V_N$, can be written as
\begin{equation}
  \label{eq:2}
  V_N= \frac{1}{2} \sum_{ij} r_i A_{ij} r_j,\quad\text{with}\quad
  A=
    \begin{pmatrix}
    \lambda_{11}&\lambda_{12}&\lambda_{13}\\
    \lambda_{12}&\lambda_{22}&\lambda_{23}\\
    \lambda_{13}&\lambda_{23}&\lambda_{33}
  \end{pmatrix}  .
\end{equation}
For the potential to be BFB, this quadratic form has to be positive
definite for $r_i\ge 0$. Then we should have the following relations
known as copositivity conditions \cite{Klimenko:1984qx,Kannike:2012pe},
\begin{align}
  \label{eq:5}
  & A_{11} \ge 0, A_{22} \ge 0, A_{33} \ge 0\, ,\nonumber\\[+2mm]
  &\overline{A}_{12}=\sqrt{A_{11}A_{22}} + A_{12} \ge 0,\quad
  \overline{A}_{13}=\sqrt{A_{11}A_{33}} + A_{13} \ge 0,\quad
  \overline{A}_{23}=\sqrt{A_{22}A_{33}} + A_{23} \ge
  0\, ,\nonumber\\[+2mm]
  &\sqrt{A_{11}A_{22}A_{33}} + A_{12}\sqrt{A_{33}}
  + A_{13}\sqrt{A_{22}}+ A_{23}\sqrt{A_{11}}
  +\sqrt{2 \overline{A}_{12}\overline{A}_{13}\overline{A}_{23}} \ge 0 .
\end{align}
This ensures that $V_N$ is BFB. For $V_{CB}$ we need two extra steps.
\subsubsection{Step 2}
This step is only necessary if at least one of the $\lambda'_{ij}$ in
Eq.~(\ref{eq:VCB}) is negative, otherwise because of Eq.~(\ref{eq:6}),
the potential along the charge breaking directions,
$V_{CB}$,  is positive definite. If at least one of the
$\lambda'_{ij}$ is negative we construct the matrices
\begin{equation}
  \label{eq:7}
  \Delta_1=
  \begin{pmatrix}
    0&\lambda'_{12}&0\\
    \lambda'_{12}&0&\lambda'_{23}\\
    0&\lambda'_{23}&0
  \end{pmatrix},\quad
  \Delta_2=
  \begin{pmatrix}
    0&0&\lambda'_{13}\\
    0&0&\lambda'_{23}\\
    \lambda'_{13}&\lambda'_{23}&0
  \end{pmatrix},\quad
  \Delta_3=
  \begin{pmatrix}
    0&\lambda'_{12}&\lambda'_{13}\\
    \lambda'_{12}&0&0\\
    \lambda'_{13}&0&0
  \end{pmatrix}.
\end{equation}
Then form the matrices
\begin{equation}
  \label{eq:8a}
  A_i=A_N + \Delta_i
\end{equation}
where $A_N$ is obtained from $V_N$. Then check the copositivity of all
$A_i$. 

\subsubsection{Step 3}
If $\lambda'_{12} \lambda'_{13}\lambda'_{23} <0$, a final step is
needed. We form the matrix \cite{Faro:2019vcd},
\begin{equation}
  \label{eq:9a}
  \Delta_4 =  \frac{1}{2}\, 
  \begin{pmatrix}\displaystyle
   \frac{\lambda'_{12}\lambda'_{13}}{\lambda'_{23}}&
    \displaystyle
    \lambda'_{12} &\displaystyle
    \lambda'_{13} \\
\displaystyle    \lambda'_{12}
& \displaystyle
 \frac{\lambda'_{12}\lambda'_{23}}{\lambda'_{13}}
&\displaystyle
\lambda'_{23}\\
\displaystyle   \lambda'_{13} & \displaystyle
\lambda'_{23}
& \displaystyle
\frac{\lambda'_{13}\lambda'_{23}}{\lambda'_{12}}
  \end{pmatrix}\, ,
\end{equation}
and construct the matrix
\begin{equation}
  \label{eq:10}
  A_4 = A_N + \Delta_4 .
\end{equation}
Now,
this matrix has to be copositive inside a tetrahedron in
the first octant and with one of
the vertices at the origin. To handle this, in Ref.~\cite{Faro:2019vcd}
the authors show that this is equivalent to finding the copositivity of
the matrix
\begin{equation}
  \label{eq:37}
  B=R^T A_4 R
\end{equation}
in the first octant, where
\begin{equation}
  \label{eq:38}
  R =
  \begin{pmatrix}
    |\lambda'_{23}| &0 &0\\
    0&|\lambda'_{13}|&0\\
    0&0&\lambda'_{12}|
  \end{pmatrix}
  \begin{pmatrix}
    0&1&1\\
    1&0&1\\
    1&1&0
  \end{pmatrix}\, .
\end{equation}

In summary, the copositivity of the matrices $A_N,A_1,A_2,A_3,B$ are
the necessary and sufficient conditions for the $U(1) \times U(1)$
potential to be BFB.

\subsection{The sufficient conditions of Ref.~\cite{Grzadkowski:2009bt}}
\label{subsec:U1U1_BFB_Grz}

We now consider the conditions from Ref.~\cite{Grzadkowski:2009bt} that are known
to be sufficient but not necessary \cite{Faro:2019vcd}. These were
derived for the case of $\Z2\times\Z2$ but our potential for $U(1)\times U(1)$ in
Eq.~(\ref{U1U1quartic}) is a particular case with, in our notation
(see Eq.~(\ref{Z2Z2quartic}) below),
\begin{equation}
  \label{eq:39}
  \lambda''_{10}=\lambda''_{11}=\lambda''_{12}=0\, .
\end{equation}
The conditions then read \cite{Grzadkowski:2009bt},
\begin{align}
  \bullet &\hskip 2mm
  \lambda_1 >0,\ \lambda_2 > 0,\ \lambda_3 >0, \label{eq:19a1}\\[+2mm]
  \bullet &\hskip 2mm
  \lambda_x > -2\sqrt{\lambda_1\lambda_2},
  \ \lambda_y > -2 \sqrt{\lambda_1\lambda_3},\
  \lambda_z >-2 \sqrt{\lambda_2\lambda_3}, \label{eq:19b1}\\[+2mm]
  \bullet &\hskip 1mm
  \left\{
    \lambda_x \sqrt{\lambda_3} +\lambda_y \sqrt{\lambda_2}
    +\lambda_z \sqrt{\lambda_1} \ge 0
  \right\}
  \cup
  \left\{
  \lambda_1
  \lambda_z^2+\lambda_2\lambda_y^2+\lambda_3\lambda_x^2
  -4\lambda_1\lambda_2\lambda_3  
  - \lambda_x\lambda_y\lambda_z < 0  
  \right\}\, ,
\label{eq:19c1}
\end{align}
where
\begin{equation}
  \label{eq:13}
  \lambda_x=\lambda_4+\text{min}(0,\lambda_7),\ \ 
  \lambda_y=\lambda_5+\text{min}(0,\lambda_8),\ \ 
  \lambda_z=\lambda_6+\text{min}(0,\lambda_9)\, .
\end{equation}

\subsection{Sufficient conditions for a lower bound}
\label{subsec:U1U1_BFB_us}

In this case we know the necessary and sufficient conditions but in many
other symmetry constrained models we do not. So we can think of a
potential that it is always lower than $V_4$ and for which the
copositivity conditions can be easily applied. This is
will be important in the following. Because of Eq.~(\ref{eq:6}),
we should have,
\begin{equation}
  \label{eq:11}
  V_{CB} \ge V_{CB}^{\rm lower}=
  r_1 r_2\ \text{min}(0,\lambda'_{12})
+ r_1 r_3\ \text{min}(0,\lambda'_{13})
+ r_2 r_3\ \text{min}(0,\lambda'_{23})
\end{equation}
and therefore
\begin{equation}
  \label{eq:4}
  V_4 \ge V_4^{\rm lower}= V_N +  V_{CB}^{\rm lower}\, .
\end{equation}
So we just have to check the copositivity of the matrix
\begin{equation}
  \label{eq:12}
  \begin{pmatrix}
    \lambda_{11}&\hat\lambda_{12}&
    \hat\lambda_{13}\\
    \hat\lambda_{12}&\lambda_{22}
    &\hat\lambda_{23}\\
    \hat\lambda_{13}
    &\hat\lambda_{23}
    &\lambda_{33}
  \end{pmatrix}\, ,
\end{equation}
where we have defined
\begin{equation}
  \label{eq:13a}
  \hat\lambda_{12}\equiv\lambda_{12}+\text{min}(0,\lambda'_{12}),\
  \hat\lambda_{13}\equiv\lambda_{13}+\text{min}(0,\lambda'_{13}),\
  \hat\lambda_{23}\equiv\lambda_{23}+\text{min}(0,\lambda'_{23}) .  
\end{equation}
These will ensure sufficient conditions for the potential to be BFB,
but they are not necessary. There will be good points in parameter
space that are discarded by this procedure. We will come to this issue
below when we compare the respective sets of points.

\section{BFB conditions in the $U(1)\times\Z2$ case}
\label{sec:BFB-U1xZ2}

The quadratic part of our $U(1)\times\Z2$ invariant potential reads,
\begin{align}
V_{\text{quartic}}=&
\lambda_1(\phi_1^\dagger\phi_1)^2
+\lambda_2(\phi_2^\dagger\phi_2)^2
+\lambda_3(\phi_3^\dagger\phi_3)^2+
\lambda_4(\phi_1^\dagger\phi_1)(\phi_2^\dagger\phi_2)
+\lambda_5(\phi_1^\dagger\phi_1)(\phi_3^\dagger\phi_3)\nonumber\\[8pt]
& 
+\lambda_6(\phi_2^\dagger\phi_2)(\phi_3^\dagger\phi_3)
+\lambda_7(\phi_1^\dagger\phi_2)(\phi_2^\dagger\phi_1)
+\lambda_8(\phi_1^\dagger\phi_3)(\phi_3^\dagger\phi_1)
+\lambda_9(\phi_2^\dagger\phi_3)(\phi_3^\dagger\phi_2)\nonumber\\[8pt]
&
+\left[\lambda''_{12}(\phi_2^\dagger\phi_3)^2
+
\text{h.c.}\right]\, ,
\label{U1Z2quartic-our}
\end{align}
satisfying
\begin{align}
  \label{eq:19}
  U(1) :&\ \phi_1 \to e^{i \theta} \phi_1,\quad \phi_2\to\phi_2,\quad
  \phi_3\to   \phi_3\, ,  \\
  \Z2 :&\ \phi_1 \to \phi_1,\quad\phi_2 \to - \phi_2,\quad
  \phi_3 \to \phi_3\, ,
\end{align}
obtained from \eqref{U'(1)_scalar} by setting $\theta'=\pi$.
In \eqref{U1Z2quartic-our}, 
``h.c.'' stands for Hermitian conjugate.
Also, we use double primes, $\lambda''_{12}$, to distinguish from the definitions in
Eq.~(\ref{eq:1}). 

\subsection{The necessary and sufficient conditions for BFB}
\label{subsec:U1Z2_BFB_NS}

The conditions for this potential to be BFB were developed by Faro and
can be found in his Master thesis \cite{Faro:2019}. In 
an adaptation of his notation\footnote{In Faro's implementation of
$U(1)\times \Z2$, $\phi_3$ is the field getting a phase.
In our notation,
this role is played by $\phi_1$.
We get from his to ours with $1 \leftrightarrow 3$.},
the non-rephasing invariant part of the potential reads
\begin{equation}
  \label{eq:14a}
  V_{U(1)\times\Z2}= \frac{1}{2}\left[\bar{\lambda}_{23}(\phi_2^\dagger\phi_3)^2
+ \text{h.c.}\right]\, .
\end{equation}
Therefore, comparing with Eq.~(\ref{U1Z2quartic-our}),
we get the relation
\begin{equation}
  \label{eq:18}
  \bar{\lambda}_{23} = 2 \lambda''_{12} .
\end{equation}
Now the BFB conditions are as in the $U(1)\times U(1)$  case doing the 3
steps mentioned there, with the substitutions
\begin{equation}
  \label{eq:21}
  \lambda_{23} \to \lambda_{23}- |\bar{\lambda}_{23}|,\quad
  \lambda'_{23} \to \lambda'_{23} + |\bar{\lambda}_{23}| .
\end{equation}

\subsection{The sufficient conditions of Ref.~\cite{Grzadkowski:2009bt}}

We now consider the sufficient conditions from Ref.~\cite{Grzadkowski:2009bt}.
They were
derived for the $\Z2\times\Z2$ case. Comparing our $U(1)\times\Z2$
potential in Eq.~(\ref{U1Z2quartic-our}) with the $\Z2\times\Z2$ case
in Eq.~(\ref{Z2Z2quartic}) we require
\begin{equation}
  \label{eq:40}
  \lambda''_{10}=\lambda''_{11}=0
\end{equation}
The conditions from Ref.~\cite{Grzadkowski:2009bt} then read,
\begin{align}
  \bullet &\hskip 2mm
  \lambda_1 >0,\ \lambda_2 > 0,\ \lambda_3 >0, \label{eq:19a2}\\[+2mm]
  \bullet &\hskip 2mm
  \lambda_x > -2\sqrt{\lambda_1\lambda_2},
  \ \lambda_y > -2 \sqrt{\lambda_1\lambda_3},\
  \lambda_z >-2 \sqrt{\lambda_2\lambda_3}, \label{eq:19b2}\\[+2mm]
  \bullet &\hskip 1mm
  \left\{
    \lambda_x \sqrt{\lambda_3} +\lambda_y \sqrt{\lambda_2}
    +\lambda_z \sqrt{\lambda_1} \ge 0
  \right\}
  \cup
  \left\{
  \lambda_1
  \lambda_z^2+\lambda_2\lambda_y^2+\lambda_3\lambda_x^2
  -4\lambda_1\lambda_2\lambda_3  
  - \lambda_x\lambda_y\lambda_z < 0  
  \right\}\, ,
\label{eq:19c2}
\end{align}
where
\begin{equation}
  \label{eq:13b}
  \lambda_x=\lambda_4+\text{min}(0,\lambda_7),\ \ 
  \lambda_y=\lambda_5+\text{min}(0,\lambda_8),\ \ 
  \lambda_z=\lambda_6+\text{min}(0,\lambda_9-2|\lambda''_{12}|)\, ,
\end{equation}
or, with the equivalence of Eq.~(\ref{eq:18}),
\begin{equation}
  \label{eq:30}
   \lambda_x=\lambda_4+\text{min}(0,\lambda_7),\ \ 
  \lambda_y=\lambda_5+\text{min}(0,\lambda_8),\ \ 
  \lambda_z=\lambda_6+\text{min}(0,\lambda_9-|\bar{\lambda}_{23}|) \, .
\end{equation}

\subsection{Sufficient conditions for a lower bound}

Although in this case there are necessary and sufficient BFB conditions,
it is instructive to
find a lower potential like in the previous case. This will serve 
to compare the set of points regarding physical observables. For the
$V_{CB}$ part, the reasoning is the same as in Eq.~(\ref{eq:11}).

Now,
for the $V_{U(1)\times \Z2}$ part, 
we note that 
\begin{equation}
(\phi_2^\dagger \phi_3)^2 + \textrm{h.c.}
=
2\, \textrm{Re}\left\{ (\phi_2^\dagger \phi_3)^2 \right\}
\geq
- 2\, \left| (\phi_2^\dagger \phi_3)^2 \right|
\geq
- 2\, |\phi_2|^2 |\phi_3|^2 = - 2 r_2 r_3\, ,
\label{simpler}
\end{equation}
where we have used the parameterization \eqref{eq:22} on the last step.
A more complicated route would be to use
\eqref{eq:22} from the start, finding
\begin{equation}
  \label{eq:23}
  V_{U(1)\times \Z2} = \bar{\lambda}_{23}\ r_2 r_3\,
f(\alpha_2,\alpha_3,\beta_2,\beta_3,\gamma)\, , 
\end{equation}
where we take  $\bar{\lambda}_{23}$ to be real but not necessarily
positive, and
\begin{align}
  \label{eq:35}
  f(\alpha_2,\alpha_3,\beta_2,\beta_3,\gamma)=&\cos ^2(\alpha_2) \cos
  ^2(\alpha_3) \cos\left[2 (\beta_2-\beta_3-\gamma)\right] +\sin ^2(\alpha_2) \sin ^2(\alpha_3) \cos (2
   \gamma)\nonumber\\[+2mm]
    &+\sin (\alpha_2) \cos
   (\alpha_2) \sin (2 \alpha_3) \cos (\beta_2-\beta_3-2 \gamma)\, .
\end{align}
Now, we can verify that we always have
\begin{equation}
  \label{eq:36}
  -1 \le f(\alpha_2,\alpha_3,\beta_2,\beta_3,\gamma) \le 1\, .
\end{equation}
Thus, using either route,
we have always
\begin{equation}
  \label{eq:24}
   V_{U(1)\times \Z2}\ge  V_{U(1)\times \Z2}^{\rm lower} = -
   |\bar{\lambda}_{23}| r_2 r_3 .
\end{equation}
Combining with Eq.~(\ref{eq:4}) we get
\begin{equation}
  \label{eq:26}
  V_4 \ge V_N + V_{CB}^{\rm lower} + V_{U(1)\times \Z2}^{\rm lower} .
\end{equation}
So we have just to look at the copositivity of the matrix
\begin{equation}
  \label{eq:12a}
  \begin{pmatrix}
    \lambda_{11}&\hat\lambda_{12}&
    \hat\lambda_{13}\\
    \hat\lambda_{12}&\lambda_{22}
    &\hat\lambda_{23}\\
    \hat\lambda_{13}
    &\hat\lambda_{23}
    &\lambda_{33}
  \end{pmatrix}\, ,
\end{equation}
where we have defined
\begin{equation}
  \label{eq:13c}
  \hat\lambda_{12}\equiv\lambda_{12}+\text{min}(0,\lambda'_{12})\, ,\
  \hat\lambda_{13}\equiv\lambda_{13}+\text{min}(0,\lambda'_{13})\, ,\
  \hat\lambda_{23}\equiv\lambda_{23}+\text{min}(0,\lambda'_{23})
  -|\bar{\lambda}_{23}|\, .  
\end{equation}
These will ensure sufficient conditions for the potential to be BFB,
but they are not necessary. There will be good points in parameter
space that are discarded by this procedure. We will come to this issue
below when we compare the respective sets of points.

\section{BFB conditions in the $\Z2\times\Z2$ case}
\label{sec:BFB-Z2xZ2}

The quadratic part of our $\Z2\times\Z2$ invariant potential reads,

\begin{align}
V_{4}=&
\lambda_1(\phi_1^\dagger\phi_1)^2
+\lambda_2(\phi_2^\dagger\phi_2)^2
+\lambda_3(\phi_3^\dagger\phi_3)^2+
\lambda_4(\phi_1^\dagger\phi_1)(\phi_2^\dagger\phi_2)
+\lambda_5(\phi_1^\dagger\phi_1)(\phi_3^\dagger\phi_3)\nonumber\\[8pt]
& 
+\lambda_6(\phi_2^\dagger\phi_2)(\phi_3^\dagger\phi_3)
+\lambda_7(\phi_1^\dagger\phi_2)(\phi_2^\dagger\phi_1)
+\lambda_8(\phi_1^\dagger\phi_3)(\phi_3^\dagger\phi_1)
+\lambda_9(\phi_2^\dagger\phi_3)(\phi_3^\dagger\phi_2)\nonumber\\[8pt]
&
+\left[\lambda''_{10}(\phi_1^\dagger\phi_2)^2 +
  \lambda''_{11}(\phi_1^\dagger\phi_3)^2 +
  \lambda''_{12}(\phi_2^\dagger\phi_3)^2
  +
\text{h.c.}\right].
\label{Z2Z2quartic}
\end{align}
satisfying
\begin{align}
 \label{eq:20}
 \Z2 :&\ \phi_1 \to -\phi_1,\quad \phi_2\to\phi_2,\quad
  \phi_3\to  \phi_3\, ,  \\
  \Z2' :&\ \phi_1 \to \phi_1,\quad\phi_2 \to -\phi_2,\quad
  \phi_3 \to \phi_3\, ,
\end{align}
which can be obtained
from Eqs.~\eqref{U(1)_scalar} and \eqref{U'(1)_scalar}
by setting $\theta=\theta'=\pi$.

The potential can be written as
\begin{equation}
  \label{eq:27}
  V_4 = V_N + V_{CB} + V_{\Z2\times\Z2} ,
\end{equation}
where $V_N$ and $V_{CB}$ are given in Eq.~(\ref{eq:VN}) and
Eq.~(\ref{eq:VCB}), respectively, and
\begin{align}
  \label{eq:28}
  V_{\Z2\times\Z2}=&\left[\lambda''_{10}(\phi_1^\dagger\phi_2)^2 +
  \lambda''_{11}(\phi_1^\dagger\phi_3)^2 +
  \lambda''_{12}(\phi_2^\dagger\phi_3)^2
  +
  \text{h.c.}\right]\nonumber\\[+2mm]
=&\ \frac{1}{2}\left[\bar{\lambda}_{12}(\phi_1^\dagger\phi_2)^2 +
  \bar{\lambda}_{13}(\phi_1^\dagger\phi_3)^2 +
  \bar{\lambda}_{23}(\phi_2^\dagger\phi_3)^2
  +
  \text{h.c.}\right]\, ,
\end{align}
where
\begin{equation}
  \label{eq:29}
  \bar{\lambda}_{12} =2 \lambda''_{10},\quad
  \bar{\lambda}_{13} =2 \lambda''_{11},\quad
  \bar{\lambda}_{23} =2 \lambda''_{12} .
\end{equation}

\subsection{The sufficient conditions of Ref.~\cite{Grzadkowski:2009bt}}
\label{sec:SuffCondsOgreid}

We now consider the sufficient conditions from Ref.~\cite{Grzadkowski:2009bt},
as implemented in Ref.~\cite{Hernandez-Sanchez:2020aop}. 
We have verified that there is a misprint in
Ref.~\cite{Hernandez-Sanchez:2020aop} when quoting
Eq.~(\ref{eq:19c3}) below, taken here from
Ref.~\cite{Grzadkowski:2009bt} (where it is correct).
We find,
\begin{align}
  \bullet &\hskip 2mm
  \lambda_1 >0,\ \lambda_2 > 0,\ \lambda_3 >0, \label{eq:19a3}\\[+2mm]
  \bullet &\hskip 2mm
  \lambda_x > -2\sqrt{\lambda_1\lambda_2},
  \ \lambda_y > -2 \sqrt{\lambda_1\lambda_3},\
  \lambda_z >-2 \sqrt{\lambda_2\lambda_3}, \label{eq:19b3}\\[+2mm]
  \bullet &\hskip 1mm
  \left\{
    \lambda_x \sqrt{\lambda_3} +\lambda_y \sqrt{\lambda_2}
    +\lambda_z \sqrt{\lambda_1} \ge 0
  \right\}
  \cup
  \left\{
  \lambda_1
  \lambda_z^2+\lambda_2\lambda_y^2+\lambda_3\lambda_x^2
  -4\lambda_1\lambda_2\lambda_3  
  - \lambda_x\lambda_y\lambda_z < 0  
  \right\}\, ,
\label{eq:19c3}
\end{align}
where
\begin{eqnarray}
  \lambda_x &=& \lambda_4+\text{min}(0,\lambda_7-2|\lambda''_{10}|)\, ,
\nonumber\\
  \lambda_y &=& \lambda_5+\text{min}(0,\lambda_8-2|\lambda''_{11}|)\, ,
\nonumber\\
  \lambda_z &=& \lambda_6+\text{min}(0,\lambda_9-2|\lambda''_{12}|)\, ,
\label{eq:13d}
\end{eqnarray}
or, with the equivalence of Eq.~(\ref{eq:29}),
\begin{eqnarray}
   \lambda_x &=& \lambda_4+\text{min}(0,\lambda_7-|\bar{\lambda}_{12}|)\, ,
\nonumber\\
  \lambda_y &=& \lambda_5+\text{min}(0,\lambda_8-|\bar{\lambda}_{13}|)\, ,
\nonumber\\
  \lambda_z &=& \lambda_6+\text{min}(0,\lambda_9-|\bar{\lambda}_{23}|)\, .
  \label{eq:30a}
\end{eqnarray}

\subsection{Sufficient conditions for a lower bound}
\label{sec:SufCondtsLowestBound}

In the $\Z2\times\Z2$, case there are no known necessary and sufficient BFB conditions.
One only has the sufficient conditions of Ref.~\cite{Grzadkowski:2009bt}
described in the previous section.
Thus,
it is interesting to
find necessary conditions from a lower potential like in the previous cases.
This will serve 
to compare the set of points regarding physical observables. For the
$V_{CB}$ part the reasoning is the same as in Eq.~(\ref{eq:11}). Now
for the $V_{\Z2\times \Z2}$ part,
we can either follow the steps in \eqref{simpler},
or use the parameterization
of Eq.~(\ref{eq:22}) to get
\begin{align}
\label{eq:31}
  V_{\Z2\times \Z2} =& \bar{\lambda}_{12}\ r_1 r_2 \cos^2(\alpha_2)
  \cos(2 \beta_2) + \bar{\lambda}_{13}\ r_1 r_3 \cos^2(\alpha_3)
  \cos\left[2 (\beta_3+\gamma)\right] \nonumber\\[+2mm]
  &+
  \bar{\lambda}_{23}\ r_2 r_3\, f(\alpha_2,\alpha_3,\beta_2,\beta_3,\gamma)\, ,
\end{align}
where we take  $\bar{\lambda}_{ij}$ to be real but not necessarily
positive.
In either case,
we have always
\begin{equation}
\label{eq:32}
V_{\Z2\times \Z2}\ge  V_{\Z2\times \Z2}^{\rm lower} =
- |\bar{\lambda}_{12}| r_1 r_2
- |\bar{\lambda}_{13}| r_1 r_3
- |\bar{\lambda}_{23}| r_2 r_3\, .
\end{equation}
Combining with Eq.~(\ref{eq:4}) we get
\begin{equation}
  \label{eq:26a}
  V_4 \ge V_N + V_{CB}^{\rm lower} + V_{\Z2\times \Z2}^{\rm lower} .
\end{equation}
So we have just to look at the copositivity of the matrix
\begin{equation}
\label{eq:33}
  \begin{pmatrix}
    \lambda_{11}&\hat\lambda_{12}&
    \hat\lambda_{13}\\
    \hat\lambda_{12}&\lambda_{22}
    &\hat\lambda_{23}\\
    \hat\lambda_{13}
    &\hat\lambda_{23}
    &\lambda_{33}
  \end{pmatrix}\, ,
\end{equation}
where we have now defined
\begin{align}
\label{eq:34}
 & \hat\lambda_{12}\equiv\lambda_{12}+\text{min}(0,\lambda'_{12})
  -|\bar{\lambda}_{12}|,\
  \hat\lambda_{13}\equiv\lambda_{13}+\text{min}(0,\lambda'_{13})
  -|\bar{\lambda}_{13}|, 
  \nonumber\\[+2mm] 
 & \hat\lambda_{23}\equiv\lambda_{23}+\text{min}(0,\lambda'_{23})
 -|\bar{\lambda}_{23}| .   
\end{align}
These will ensure sufficient conditions for the potential to be BFB,
but they are not necessary. There will be good points in parameter
space that are discarded by this procedure. We will come to this issue
below when we compare the respective sets of points.

\section{\label{sec:scalar}Setup of the models: scalar sector}

To be able to compare the phenomenological impact
of the various BFB conditions, we generalized our
previous numerical
code \cite{Fontes:2014xva,Fontes:2017zfn,Florentino:2021ybj,Boto:2021qgu}
to the symmetry constrained potentials we consider here; namely,
$U(1)\times U(1)$, $U(1)\times\Z2$, and $\Z2\times\Z2$. Comparing
Eq.~(\ref{Z2Z2quartic}) with Eq.~(\ref{U1Z2quartic-our}) and
Eq.~(\ref{U1U1quartic}) we see that the first two can be obtained
form the last by setting some or all of the couplings $\lambda''_{ij}$
to zero. To get all the necessary couplings we implemented the case
$\Z2\times\Z2$ in
\texttt{FeynMaster} \cite{Fontes:2019wqh,Fontes:2021iue}, the others
follow from the argument above.

As we want to define the relations of the couplings to masses and
angles, we have to go back and consider the full potential
\begin{equation}\label{potential}
    V=V_{2}+V_{4},
\end{equation}
where the quartic part, $V_4$, is given in
Eqs.~(\ref{U1U1quartic}),~(\ref{U1Z2quartic-our})
and~(\ref{Z2Z2quartic}), depending on which case we consider, and
the quadratic part is,
\begin{equation}
    V_{2}=m_{11}^2\phi_1^\dagger\phi_1+m_{22}^2\phi_2^\dagger\phi_2+m_{33}^2\phi_3^\dagger\phi_3 +\left[m_{12}^2(\phi_1^\dagger\phi_2)
+m_{13}^2(\phi_1^\dagger\phi_3)
+m_{23}^2(\phi_2^\dagger\phi_3)+\text{h.c.}\right] ,
\end{equation}
where we also include terms, $m_{12}^2$, $m_{13}^2$ and $m_{23}^2$,
that break the symmetry softly. In our study we consider that the
potentially complex parameters, $m^2_{12},m^2_{13},m^2_{23}$ and 
$\lambda''_{10},\lambda''_{11},\lambda''_{12}$ are taken real.

After spontaneous symmetry breaking (SSB), the three doublets can be
parameterized in terms of its component fields as: 
 \begin{equation}
     \phi_i=\begin{pmatrix} w_k^\dagger \\ (v_i+x_i\,+\,i\,z_i)/\sqrt{2}\end{pmatrix} \,\,\qquad (i=1,2,3)\label{fielddefinitions}\, ,
 \end{equation}
where $v_i/\sqrt{2}$ corresponds to the vacuum expectation value
(vev) for the neutral component of $\phi_i$.
It is assumed that the
scalar sector of the
model explicitly and spontaneously conserves
CP.\footnote{Strictly speaking,
it is not advisable to assume a real scalar sector while allowing the
Yukawa couplings to carry the phase necessary for the
CKM matrix.
This is also a problem with the so-called real 2HDM
\cite{Fontes:2021znm}.
One can take the view that the complex terms and their counterterms
in the scalar sector exist, with the former set to zero.
}

That is, all the parameters in the scalar potential are real and
the vevs $v_1$, $v_2$, and $v_3$ are also real.
With this assumption, the scalar potential 
contains at most, eighteen parameters. The vevs can be parameterized as follows:
\begin{equation}\label{3hdmvevs}
     v_1=v \cos \beta_1 \cos \beta_2\,,\qquad v_2=v \sin \beta_1 \cos \beta_2\, ,\qquad v_3=v \sin \beta_2,
\end{equation}
leading to the Higgs basis  \cite{Georgi:1978ri,Donoghue:1978cj,Botella:1994cs}
to be obtained by the following rotation,
\begin{equation}\label{higgsbasisZ3}
     \begin{pmatrix} H_0 \\ R_1 \\ R_2 \end{pmatrix}
     =
     \mathcal{O}_\beta
     \begin{pmatrix} x_1 \\ x_2 \\ x_3 \end{pmatrix}
     =
     \begin{pmatrix} \cos\beta_2 \cos\beta_1 & \cos\beta_2 \sin\beta_1 & \sin\beta_2 \\ -\sin\beta_1 & \cos\beta_1 & 0 \\ -\cos\beta_1 \sin\beta_2 & -\sin\beta_1 \sin\beta_2 & \cos\beta_2\end{pmatrix}
     \begin{pmatrix} x_1 \\ x_2 \\ x_3 \end{pmatrix} .
\end{equation}
The scalar kinetic Lagrangian is written as
\begin{equation}\label{kinetic3hdm}
    \mathscr{L}_{\text{kin}}=\sum_{k=1}^{n=3}|D_\mu\phi_k|^2 ,
\end{equation}
and contains the terms relevant to the propagators and trilinear couplings of the scalars and gauge bosons.

 We can now define orthogonal matrices which diagonalize the
 squared-mass matrices present in the CP-even scalar, CP-odd scalar
 and charged scalar sectors. These are the transformations that take
 us to the physical basis, with states possessing well-defined
 masses. Following Ref.~\cite{Das:2019yad,Boto:2021}, the twelve
 quartic couplings for the $\Z2\times\Z2$ can be exchanged for seven
 physical masses (three  
CP-even scalars, two CP-odd scalars and two pairs of charged scalars)
and five mixing angles. For the case of $U(1)\times\Z2$, we have only
10 $\lambda$'s and therefore we can also solve for two of the soft
masses.
Finally, in the case of $U1)\times U(1)$ one has only 9
$\lambda$'s, and one can also solve for all the soft masses.
We give all the explicit expressions in Appendix~\ref{app:lambdas}.

The mass terms in the neutral scalar sector
can be extracted through the following rotation,  
\begin{equation}\label{CPevenDiag}
     \begin{pmatrix} h_1 \\ h_2 \\ h_3 \end{pmatrix}
=\mathcal{O}_\alpha \begin{pmatrix} x_1 \\ x_2 \\ x_3 \end{pmatrix},
\end{equation}
where we take $h_1 \equiv h_{125}$ to be the 125GeV Higgs particle
found at LHC.
The form chosen for $\mathcal{O}_\alpha$ is
\begin{equation}\label{matrixR}
\textbf{R}\equiv\mathcal{O}_\alpha=\mathcal{R}_3.\mathcal{R}_2.\mathcal{R}_1 ,
\end{equation}
where
\begin{equation}
\mathcal{R}_1=\begin{pmatrix}\cai & \sai & 0\\ -\sai & \cai & 0 \\ 0 & 0 & 1  \end{pmatrix}\,,\quad     \mathcal{R}_2=\begin{pmatrix}\caii & 0 & \saii \\ 0 & 1 & 0 \\ -\saii & 0 & \caii  \end{pmatrix}\,,\quad     \mathcal{R}_3=\begin{pmatrix}1 & 0 & 0\\ 0 & \caiii & \saiii \\ 0 & -\saiii & \caiii  \end{pmatrix}\,.\quad
\end{equation}

For the CP-odd scalar sector, the physical basis is
chosen as $\begin{pmatrix}G^0 & A_1 & A_2\end{pmatrix}^T$
and the transformation to be
\begin{equation}\label{CPoddDiag}
\begin{pmatrix} G^0 \\ A_1 \\ A_2 \end{pmatrix}
=
\mathcal{O}_{\gamma_1} \mathcal{O}_\beta
\begin{pmatrix} z_1 \\ z_2 \\ z_3 \end{pmatrix} ,
\end{equation}
where
\begin{equation}
\mathcal{O}_{\gamma_1}
=
\begin{pmatrix}1  & 0 & 0\\ 0& \cgi & -\sgi \\ 0 & \sgi & \cgi \end{pmatrix}
\label{ogamma1}
\end{equation}
is defined in order to diagonalize the 2x2 submatrix
that remains non-diagonal in the Higgs basis.
For later use, we define the matrix $\textbf{P}$
as the combination
\begin{equation}\label{matrixP}
    \textbf{P}\equiv\mathcal{O}_{\gamma_1} \mathcal{O}_\beta.
\end{equation}

For the charged scalar sector, the physical basis is 
$\begin{pmatrix}G^+ & H_1^+ & H_2^+\end{pmatrix}^T$
and the transformation is
\begin{equation}\label{ChargedDiag}
\begin{pmatrix} G^+ \\ H_1^+ \\ H_2^+ \end{pmatrix}
=\mathcal{O}_{\gamma_2} \mathcal{O}_\beta
\begin{pmatrix} w_1^\dagger \\ w_2^\dagger \\ w_3^\dagger \end{pmatrix},
\end{equation}
where
\begin{equation}
\mathcal{O}_{\gamma_2}= \begin{pmatrix}1  & 0 & 0\\ 0& \cgii & -\sgii \\ 0 & \sgii & \cgii \end{pmatrix} .
\label{ogamma2} 
\end{equation}
We write the masses of $H_1^+$ and $H_2^+$
as $m_{H_1^\pm}$ and $m_{H_2^\pm}$, respectively.
The matrix $\textbf{Q}$ is defined as the combination 
\begin{equation}\label{matrixQ}
\textbf{Q}\equiv\mathcal{O}_{\gamma_2} \mathcal{O}_\beta .
\end{equation}
The matrix $\textbf{Q}$ is relevant for the calculation
of the oblique parameters,
which we relegate to Appendix~\ref{app:STU},
following the analysis of \cite{Grimus:2007if}.

For completeness, we also include in Appendix~\ref{app:Unitarity}
the perturbative unitarity constraints,
following \cite{Bento:2017eti,Bento:2022vsb}.

Considering that the states in the physical basis have well-defined masses,
we can obtain relations between the set 
\begin{align}
&&\left\{v_1,v_2,v_3,m_{h_1},m_{h_2},m_{h_3},m_{A_1},m_{A_2},
m_{H_1^\pm},m_{H_2^\pm},\alpha_1,\alpha_2,\alpha_3,
\gamma_1,\gamma_2,m^2_{12},m^2_{13},m^2_{23}\right\}
,\label{setphysical}\\[8pt] 
&&\quad 
v_1=v \cos\beta_1 \cos\beta_2\,,\quad\,v_2=v \sin\beta_1 \cos\beta_2\,,\quad\,v_3=v \sin\beta_2 ,
\end{align}
and the parameters of the potential\footnote{As mentioned above,
for the $U(1)\times U(1)$ and $U(1)\times\Z2$ cases,
since we have less parameters, some or all of
the soft mass squared terms can also be solved for,
as shown explicitly in Appendix~\ref{app:lambdas}.}
as shown in Ref.~\cite{Das:2019yad,Boto:2021}.

\section{\label{sec:Yukawas}Setup of the models: Yukawa interactions}

The most general quark Yukawa Lagrangian of the 3HDM may be written as
\be
\mathcal{L}_\mathrm{Y} =
- \bar q_L \left[
\left( \Gamma_1 \phi_1 + \Gamma_2 \phi_2
+ \Gamma_3 \phi_3\right) n_R
+
\left( \Delta_1 \tilde \phi_1 + \Delta_2 \tilde \phi_2
+ \Delta_3 \tilde \phi_3 \right) p_R
\right]
+ \mathrm{h.c.},
\label{yuk-q}
\ee
where $\tilde \phi_k \equiv i \tau_2 \phi_k^\ast$,
while $q_L$, $n_R$, and $p_R$
are vectors~\footnote{These vectors are written in a weak basis;
not in the mass basis. For massless neutrinos,
we can take the leptons already in the mass basis.}
in the respective three-dimensional flavour
vector space of left-handed quark doublets,
right-handed down-type quarks,
and right-handed up-type quarks.

Ignoring neutrino masses, the leptonic
Yukawa Lagrangian of the 3HDM may be similarly written as
\be
\mathcal{L}_\mathrm{Y} =
- \bar L_L \left[
\left( \Pi_1 \phi_1 + \Pi_2 \phi_2
+ \Pi_3 \phi_3\right) \ell_R
\right]
+ \mathrm{h.c.},
\label{yuk-lep}
\ee
where $L_L$ and $\ell_R$
are vectors in the respective three-dimensional flavour
vector space of left-handed leptonic doublets
and right-handed charged leptons.
The $3 \times 3$ matrices
$\Gamma_k$,
$\Delta_k$,
and $\Pi_k$
contain the complex Yukawa couplings
to the right-handed down-type quarks, up-type quarks,
and charged leptons,
respectively.

As is well known,
unless protected by a symmetry,
the Higgs-fermion Yukawa couplings
lead to Higgs-mediated flavor-changing neutral couplings (FCNC)
at a level incompatible with experimental observations.
FCNC can be removed by making 
the Yukawa coupling matrices to fermions of a given electric charge
proportional:
\be
\Gamma_1 \propto \Gamma_2 \propto \Gamma_3\, ,
\ \ \ 
\Delta_1 \propto \Delta_2 \propto \Delta_3\, ,
\ \ \ 
\Pi_1 \propto \Pi_2 \propto \Pi_3\, .
\label{propto}
\ee
It has been shown that,
in a general NHDM,
Eqs.~\eqref{propto}
remain true (thus removing FCNCs) under the renormalization
group running if and only if there is a basis for the Higgs doublets
in which all the fermions of a given electric charge couple to
only one Higgs doublet \cite{Ferreira:2010xe}.
This can be imposed in the 2HDM through
a $\Z2$ symmetry \cite{Glashow:1976nt,Paschos:1976ay},
leading to four types of
models.
For $N\geq 3$ there are five possible choices \cite{Ferreira:2010xe},
which \cite{Yagyu:2016whx} dubbed
Types I, II, X, Y, and Z,
as 
\begin{eqnarray}
\textrm{Type-I:}
&&
\phi_u=\phi_d=\phi_e\, ,
\nonumber\\
\textrm{Type-II:}
&&
\phi_u \neq \phi_d=\phi_e\, ,
\nonumber\\
\textrm{Type-X:}
&&
\phi_u=\phi_d \neq \phi_e\, ,
\nonumber\\
\textrm{Type-Y:}
&&
\phi_u=\phi_e \neq \phi_d\, ,
\nonumber\\
\textrm{Type-Z:}
&&
\phi_u\neq \phi_d;\ \phi_d\neq \phi_e,\ \phi_e\neq \phi_u\, ,
\label{Types}
\end{eqnarray}
with $\phi_{u,d,e}$ being the single scalar fields that couple
exclusively to the up-type quarks, down-type quarks, and charged
leptons, respectively.  

We wish to see how these choices can be implemented in
the $U(1)\times U'(1)$ symmetric 3HDM.
(In this section,
we briefly change the notation from
$U(1)\times U(1)$, $U(1)\times\Z2$, and $\Z2\times\Z2$,
into
$U(1)\times U'(1)$, $U(1)\times \mathbbm{Z}'_2$, and
$\Z2\times \mathbbm{Z}'_2$,
respectively.)
Without loss of generality,
we can choose $\phi_u=\phi_3$,
with the scalar fields transforming under
$U(1)$ and $U'(1)$, respectively, as
\begin{align}
U(1) &:   &   \phi_1 &\rightarrow e^{i \theta}\phi_1
&   \phi_2 &\rightarrow \phi_2
&   \phi_3 &\rightarrow \phi_3\, ,
\label{U(1)_scalar-2}
\\
U'(1) &:   &   \phi_1 &\rightarrow \phi_1
&   \phi_2 &\rightarrow e^{i \theta'}\phi_2
&   \phi_3 &\rightarrow \phi_3\, .
\label{U'(1)_scalar-2}
\end{align}
We choose three fields to remain invariant under the two groups:
\be
U(1)\ \textrm{and} \ U'(1):
q_L \rightarrow q_L\, ,
\ \ 
p_R \rightarrow p_R\, ,
\ \
L_L \rightarrow L_L\, ,
\label{U(1)ANDU'(1)}
\ee
under both $U(1)$ and $U'(1)$.
Eqs.~\eqref{U(1)_scalar-2}, \eqref{U'(1)_scalar-2}, and
\eqref{U(1)ANDU'(1)} ensure that $\phi_3=\phi_u$.
The various types can now be implemented by
choosing the other fields to transform as in
Table~\ref{tab:Types}.
\begin{table}[htb]
   \centering
   \begin{tabular}{|c||c|c|c|c|c|c|}\hline
     &$\phi_1$&$\phi_2$&$\phi_3$&$n_R$&$\ell_R$&$\phi_u$
     $\phi_d$ $\phi_{\ell}$\\\hline\hline
    Type-I &$(e^{\, i \theta},\hspace{7mm})$ &$(\hspace{7mm},e^{\, i \theta'})$ &$(\hspace{7mm},\hspace{7mm})$
    &$(\hspace{7mm},\hspace{7mm})$ &$(\hspace{7mm},\hspace{7mm})$
     &$\phi_3$ $\phi_3$ $\phi_3$\\*[1mm]
     Type-II &$(e^{\, i \theta},\hspace{7mm})$ &$(\hspace{7mm},e^{\, i \theta'})$ &$(\hspace{7mm},\hspace{7mm})$
     &$(\hspace{7mm},e^{-i \theta'})$ & $(\hspace{7mm},e^{-i \theta'})$
     & $\phi_3$ $\phi_2$ $\phi_2$\\*[1mm]
     Type-X &$(e^{\, i \theta},\hspace{7mm})$ &$(\hspace{7mm},e^{\, i \theta'})$ & $(\hspace{7mm},\hspace{7mm})$
     &$(\hspace{7mm},\hspace{7mm})$ & $(\hspace{7mm},e^{-i \theta'})$
     & $\phi_3$ $\phi_3$ $\phi_2$\\*[1mm]
     Type-Y & $(e^{\, i \theta},\hspace{7mm})$&$(\hspace{7mm},e^{\, i \theta'})$ &$(\hspace{7mm},\hspace{7mm})$
     &$(\hspace{7mm},e^{-i \theta'})$ &$(\hspace{7mm},\hspace{7mm})$
     & $\phi_3$ $\phi_2$ $\phi_3$\\*[1mm]
     Type-Z &$(e^{\, i \theta},\hspace{7mm})$ &$(\hspace{7mm},e^{\, i \theta'})$ & $(\hspace{7mm},\hspace{7mm})$
     &$(\hspace{7mm},e^{-i \theta'})$ &$(e^{-i \theta},\hspace{7mm})$
     & $\phi_3$ $\phi_2$ $\phi_1$\\\hline
   \end{tabular}
   \caption{All possible models with natural flavour conservation.
   The transformation
     properties under $U(1)\times U'(1)$ are indicated by
     $(\ ,\ )$. For instance $(\ \ ,e^{i \theta'})$ indicates that the field is
     invariant under the first $U(1)$ but transforms as
     $\psi \rightarrow e^{i \theta'} \psi$ under $U'(1)$.
     For $U(1)\times \mathbbm{Z}'_2$ do
     $e^{\pm i \theta'} \rightarrow -$, and
     for $\Z2\times \mathbbm{Z}'_2$ do
     $e^{\pm i \theta}, e^{\pm i \theta'} \rightarrow -$.
     }
     \label{tab:Types}
 \end{table}


The transformations of the fields under
$U(1)\times \mathbbm{Z}'_2$ are obtained from Table~\ref{tab:Types}
by changing $e^{\pm i \theta'} \rightarrow -$.
Similarly,
The transformations of the fields under
$\Z2\times \mathbbm{Z}'_2$ are obtained from Table~\ref{tab:Types}
by changing both $e^{\pm i \theta} \rightarrow -$ and
$e^{\pm i \theta'} \rightarrow -$.

We treat in this main text in detail the \textbf{Type-I} models.
The remaining Types are relegated to Appendix~\ref{app:Types}.
For this case we assume that under the group
all the fermion fields are unaffected. Therefore they can only couple
to $\phi_3$.
When taking into account the restrictions imposed by the symmetry, the
Yukawa couplings to fermions can be written in a compact form. For the
couplings of neutral Higgs to fermions, 
\begin{equation}\label{eq:couplingNeutralFerm}
    \mathscr{L}_{\rm Y}\ni -\frac{m_f}{v}\bar{f}(a^f_j+i\, b^f_j\gamma_5)fh_j ,
\end{equation}
where we group the physical Higgs fields in a vector, as
$h_j\equiv(h_1,h_2,h_3,A_1,A_2)_j$. We have
\begin{align}
a_j^f \to&
\frac{\textbf{R}_{j,3}}{\hat{v_3}},
\qquad\qquad j=1,2,3\qquad \text{for all leptons} ,\nonumber\\[2pt]
b_j^f \to&
\frac{\textbf{P}_{j-2,3}}{\hat{v_3}},
\qquad\quad j=4,5\quad\qquad \text{for all leptons} ,\nonumber\\[2pt]
a_j^f \to&
\frac{\textbf{R}_{j,3}}{\hat{v_3}},
\qquad\qquad j=1,2,3\qquad \text{for all up quarks} ,\nonumber\\[2pt]
b_j^f \to&
-\frac{\textbf{P}_{j-2,3}}{\hat{v_3}},
\quad\quad j=4,5\quad\qquad \text{for all up quarks} ,\nonumber\\[2pt]
a_j^f \to&
\frac{\textbf{R}_{j,3}}{\hat{v_3}},
\qquad\qquad j=1,2,3\qquad \text{for all down quarks} ,\nonumber\\[2pt]
b_j^f \to&
\frac{\textbf{P}_{j-2,3}}{\hat{v_3}},
\qquad\quad j=4,5\quad\qquad \text{for all down quarks} ,
\label{eq:coeffNeutralFerm-Type-I}
\end{align}
The couplings of the charged Higgs, $H_1^\pm$ and $H_2^\pm$, to fermions can be expressed as
\begin{eqnarray}
\mathscr{L}_{\rm Y} &\ni& \frac{\sqrt{2}}{v}
\bar{\psi}_{d_i}
\left[m_{\psi_{d_i}} V_{ji}^\ast\, \eta_k^L P_L
+ m_{\psi_{u_j}} V_{ji}^\ast\, \eta_k^R P_R\right] \psi_{u_j} H_k^-
\nonumber\\
&&
+ \frac{\sqrt{2}}{v}\bar{\psi}_{u_i}
\left[m_{\psi_{d_j}} V_{ij}\, \eta_k^L P_R 
+ m_{\psi_{u_i}} V_{ij}\, \eta_k^R P_L \right] \psi_{d_j} H_k^+ ,
\label{eq:couplingChargedFerm}
\end{eqnarray}
where $(\psi_{u_i},\psi_{d_i})$ is $(u_i,d_i)$ for
quarks\footnote{Here, the up-type quarks $u$ and
down-type quarks $d$ are already written in the mass basis.}
or $(\nu_i,\ell_i)$ for leptons.
For quarks, $V$ is the CKM matrix, while for leptons,
$V_{ij}=\delta_{ij}$ since we are considering massless neutrinos. The
couplings are,
\begin{equation}
\label{eq:coeffChargedFerm-Type-I}  
\eta_k^{\ell\,L}=-\frac{\textbf{Q}_{k+1,3}}{\hat{v_3}}\,,\quad\eta_k^{\ell\,R}=
0\,,\quad\eta_k^{q\,L} =-\frac{\textbf{Q}_{k+1,3}}{\hat{v_3}}\,,
\quad\eta_k^{q\,R}=\frac{\textbf{Q}_{k+1,3}}{\hat{v_3}}\,,\quad
\text{k=1,2}\, .
\end{equation}

\section{Kappas}
\label{sec:kappas}

We found that it is useful to select points that are already close to
the LHC constraints, using the $\kappa$'s formalism. We require them
to be within 3$\sigma$ of the LHC data \cite{ATLAS-CONF-2018-031}.
This is used to generate an initial set of points, to be
improved on below.
We
list below the expressions for the kappas for the various types. For
all types we have
\begin{equation}
  \label{eq:1a}
  \kappa_W=\cos(\alpha_2) \cos(\alpha_1-\beta_1) \cos(\beta_2) +
  \sin(\alpha_2) \sin(\beta_2)\, ,
\end{equation}
which gives $\kappa_W=1$ when $\alpha_1=\beta_1$ and
$\alpha_2=\beta_2$. 

\subsection{Type-I}
We have
\begin{equation}
  \kappa_U= \frac{\sin(\alpha_2)}{\sin(\beta_2)},
\quad
  \kappa_D= \frac{\sin(\alpha_2)}{\sin(\beta_2)},
  \quad
  \kappa_L=  \frac{\sin(\alpha_2)}{\sin(\beta_2)}.
\end{equation}

\subsection{Type-II}
We have
\begin{equation}
  \kappa_U= \frac{\sin(\alpha_2)}{\sin(\beta_2)},
\quad
  \kappa_D=\frac{\sin(\alpha_1) \cos(\alpha_2)}{\sin(\beta_1) \cos(\beta_2)},
  \quad
  \kappa_L=  \frac{\sin(\alpha_1) \cos(\alpha_2)}{\sin(\beta_1) \cos(\beta_2)}.
\end{equation}

\subsection{Type-X}
We have
\begin{equation}
  \kappa_U= \frac{\sin(\alpha_2)}{\sin(\beta_2)},
\quad
  \kappa_D=\frac{\sin(\alpha_2)}{\sin(\beta_2)},
  \quad
  \kappa_L= \frac{\sin(\alpha_1) \cos(\alpha_2)}{\sin(\beta_1) \cos(\beta_2)}. 
\end{equation}

\subsection{Type-Y}
We have
\begin{equation}
  \kappa_U= \frac{\sin(\alpha_2)}{\sin(\beta_2)},
\quad
  \kappa_D=\frac{\sin(\alpha_1) \cos(\alpha_2)}{\sin(\beta_1) \cos(\beta_2)},
  \quad
  \kappa_L=  \frac{\sin(\alpha_2)}{\sin(\beta_2)}.
\end{equation}

\subsection{Type-Z}

We have
\begin{equation}
  \kappa_U= \frac{\sin(\alpha_2)}{\sin(\beta_2)},
\quad
  \kappa_D=\frac{\sin(\alpha_1) \cos(\alpha_2)}{\sin(\beta_1) \cos(\beta_2)},
  \quad
  \kappa_L= \frac{\cos(\alpha_1) \cos(\alpha_2)}{\cos(\beta_1) \cos(\beta_2)}.
\end{equation}

\section{Scan strategy and Constraints}
\label{ref:scan}

\subsection{The scan}

For each of the three symmetry constrained 3HDM, we built a dedicated
code, which is an extension of our previous
codes~\cite{Fontes:2014xva,Florentino:2021ybj,Boto:2021qgu}.
We performed an extensive scan of the parameter space in \eq{setphysical}.
Our fixed inputs are $v = 246\,\text{GeV}$ and $m_{h1} = 125\,\text{GeV}$.
We then took random values in the ranges:
\begin{align}
&\alpha_1,\, \alpha_2,\, \alpha_3,\, \gamma_1,\, \gamma_2\, \in \left[-\frac{\pi}{2},\frac{\pi}{2}\right];\qquad \tan{\beta_1},\,\tan{\beta_2}\,\in \left[0,10\right];
\nonumber\\[8pt]
&m_{H_1}\equiv m_{h_2},\, m_{H_2}\equiv m_{h_3}\,
\in \left[125,1000\right]\,\text{GeV};\\[8pt]
&
m_{A_1},\,m_{A_2}\,m_{H_1^\pm},\,m_{H_2^\pm}\,
\in \left[100,1000\right]\,\text{GeV};\\[8pt]
&
m^2_{12},m^2_{13},m^2_{23} \in  \left[\pm 10^{-1},\pm 10^{7}\right]\,
\text{GeV}^2\, ,
\label{eq:scanparameters}
\end{align}
where the last expression applies only to the soft
masses that are not obtained as derived quantities (see the expressions
in Appendix~\ref{app:lambdas}).
These parameter ranges will be used in all scans and figures
presented below, except where noted otherwise.
The lower limits chosen for the masses satisfy the
constraints listed in
Ref.~\cite{Aranda:2019vda}.

When studying 3HDM, it was
noted~\cite{Das:2019yad,Boto:2021qgu,Chakraborti:2021bpy} that in
order to be able to generate good points in an easy way one should not be far away
from alignment, defined as the situation where the lightest Higgs
scalar has the SM couplings. It was shown in Ref.~\cite{Das:2019yad}
that this corresponds to the case when
\begin{equation}
  \label{eq:43}
  \alpha_1=\beta_1,\quad \alpha_2=\beta_2, 
\end{equation}
with the remaining parameters allowed to be free, although subject to
the constraints below. It turns out that for $\Z3$ 3HDM~\cite{Boto:2021qgu}, this
constraint alone is not enough to generate a reasonably large set of
good points starting from a completely unconstrained scan as in
Eq.~(\ref{eq:scanparameters}). Ref.~\cite{Chakraborti:2021bpy}
noticed a quite remarkable situation. If, besides the alignment of
Eq.~(\ref{eq:43}), one also requires
\begin{equation}
  \label{eq:44}
  \gamma_1=\gamma_2=-\alpha_3,\quad m_{H_1}=m_{A_1}=m_{H_1^\pm},\quad
  m_{H_2}=m_{A_2}=m_{H_2^\pm} ,
\end{equation}
then the potentials of Eq.~(\ref{U1U1quartic}),
(\ref{U1Z2quartic-our}) and (\ref{Z2Z2quartic}) all collapse into a
very symmetric form,
\begin{equation}
  \label{eq:45}
  V_{\rm Sym\, Lim} = \lambda_{\rm SM} \left[ (\phi_1^\dagger \phi_1) +
    (\phi_2^\dagger \phi_2) +(\phi_3^\dagger \phi_3) \right]^2\, ,
\end{equation}
with
\begin{equation}
  \label{eq:46}
  \lambda_{\rm SM}= \frac{m_h^2}{2 v^2} ,
\end{equation}
being the SM quartic Higgs coupling. This requires that, for the
conditions in Eq.~(\ref{eq:43}) and Eq.~(\ref{eq:44}), we have
\begin{equation}
  \label{eq:47}
  \lambda_1=\lambda_2=\lambda_3=\lambda_{\rm SM},\ \ \ 
  \lambda_4=\lambda_5=\lambda_6=2\lambda_{\rm SM},
\end{equation}
with all other $\lambda'$s vanishing. Imposing the validity of Eq.~(\ref{eq:43})
and Eq.~(\ref{eq:44}) also implies that the soft masses can be explicitly
solved as,
\begin{align}
  \label{eq:softs-sym}
  m^2_{12}=&c_{\beta_{1}}^2 c_{\gamma_{2}} s_{\beta_{2}} s_{\gamma_{2}}
   \left(m_{H^{+}_{1}}^2-m_{H^{+}_{2}}^2\right)+c_{\beta_{1}} s_{\beta_{1}}
   \left[s_{\beta_{2}}^2 \left(c_{\gamma_{2}}^2 m_{H^{+}_{2}}^2+m_{H^{+}_{1}}^2
   s_{\gamma_{2}}^2\right)-c_{\gamma_{2}}^2 m_{H^{+}_{1}}^2-m_{H^{+}_{2}}^2
 s_{\gamma_{2}}^2\right]\nonumber\\
&+c_{\gamma_{2}} s_{\beta_{1}}^2 s_{\beta_{2}} s_{\gamma_{2}}
 \left(m_{H^{+}_{2}}^2-m_{H^{+}_{1}}^2\right)\\[+3mm]
 m^2_{13}=&-c_{\beta_{2}} \left[c_{\beta_{1}} s_{\beta_{2}} \left(c_{\gamma_{2}}^2
   m_{H^{+}_{2}}^2+m_{H^{+}_{1}}^2 s_{\gamma_{2}}^2\right)-c_{\gamma_{2}} s_{\beta_{1}}
   s_{\gamma_{2}} \left(m_{H^{+}_{1}}^2-m_{H^{+}_{2}}^2\right)\right]\\[+3mm]
 m^2_{23}=& -c_{\beta_{2}} \left[c_{\beta_{1}} c_{\gamma_{2}} s_{\gamma_{2}}
   \left(m_{H^{+}_{1}}^2-m_{H^{+}_{2}}^2\right)+s_{\beta_{1}} s_{\beta_{2}}
   \left(c_{\gamma_{2}}^2 m_{H^{+}_{2}}^2+m_{H^{+}_{1}}^2 s_{\gamma_{2}}^2\right)\right]
 \end{align}
We have verified that this works
not only for the case of the symmetry constrained $\Z3$ of Refs.~\cite{Boto:2021qgu,Chakraborti:2021bpy},
but also for the
case of $U(1)\times U(1)$, $U(1)\times\Z2$ and $\Z2\times\Z2$. Now it
is easy to understand that all such points are good points. Due to
alignment, the LHC results on the $h_{125}$ are easily obeyed, while
the perturbativity unitarity, STU and the other constraints are
automatically obeyed. In fact we are quite close to the SM.

In studying the 3HDM with $\Z3$ we found \cite{Boto:2021qgu} that we
could go away from the conditions of Eq.~(\ref{eq:43}) and Eq.~(\ref{eq:44}) 
by a given percentage (10\%, 20\%, 50\%) and enhance the possibility
of some signals, while at the same time being able to generate enough
points. We checked this again for the three symmetry constrained 3HDM
we study here. Nevertheless, for the case of $U(1)\times\Z2$,  and
specially for $U(1)\times U(1)$, we could generate a large set of
points just implementing a percentage of 50\% around
Eq.~(\ref{eq:43}).
That is
\begin{equation}
  \label{Al-1}
  \frac{\alpha_1}{\beta_1},\ 
    \frac{\alpha_2}{\beta_2}\, \in\, [0.5,1.5]\, ,
    \ \ \ \textbf{(Al-1)}
\end{equation}
and not imposing the conditions in Eq.~(\ref{eq:44}).
This first (and less stringent) alignment condition will be
denoted by ``Al-1'' below.
The second, more stringent alignment condition,
combines Al-1 with six new conditions,
\begin{equation}
  \label{Al-2}
  \frac{\alpha_1}{\beta_1},\ 
  \frac{\alpha_2}{\beta_2},\ 
  \frac{\gamma_2}{\gamma_1},\ 
  \frac{-\alpha_3}{\gamma_1},\ 
  \frac{m_{A_1}}{m_{H_1}},\ 
  \frac{m_{H_1^\pm}}{m_{H_1}},\ 
  \frac{m_{A_2}}{m_{H_2}},\ 
    \frac{m_{H_2^\pm}}{m_{H_2}}\, \in\, [0.5,1.5]\, ,
    \ \ \ \textbf{(Al-2)}
\end{equation}
and will be denoted by ``Al-2'' below.
For the soft masses, in the
cases where they are independent parameters, we also use the same
approach as in Eq.~(\ref{Al-2}) with respect to Eq.~(\ref{eq:softs-sym}).

\subsection{\label{sect:constraints}Constraints on the parameter space}

In this section we study the constraints that must be applied to the
model parameters in order to ensure consistency. They are both
theoretical (consistency of the model) and experimental, as we
describe below.

\subsubsection{BFB Conditions}

We start with the BFB conditions. As explained in 
sections~\ref{sec:BFB-U1xU1}, \ref{sec:BFB-U1xZ2} and
\ref{sec:BFB-Z2xZ2}, we do not have necessary and sufficient
conditions for the $\Z2\times\Z2$ case, only for the $U(1)\times U(1)$ and
$U(1)\times\Z2$ cases. Therefore for the $\Z2\times\Z2$ case we use only
the sufficient conditions of Section~\ref{sec:SuffCondsOgreid} or of
Section~\ref{sec:SufCondtsLowestBound}. One of the main results of
this study is the comparison, for the case when we have a necessary and
sufficient condition, between the set of points that pass this
condition, with those that are eliminated by more restrictive sufficient
conditions. We will do this 
study below for the case of the $U(1)\times U(1)$ symmetric 3HDM.

 \subsubsection{Perturbative Unitarity}

In order to determine the tree-level unitarity constraints, we use
the algorithm presented in \cite{Bento:2017eti},
as described in Appendix~\ref{app:Unitarity}.

\subsubsection{Oblique parameters $STU$}

In order to discuss the effect of the $S,T,U$ parameters,
we use the expressions in \cite{Grimus:2007if} and the
experimental summary in \cite{Baak:2014ora}.
We explain in Appendix~\ref{app:STU} how to implement
this in our class of models.

\subsubsection{Perturbative Yukawa couplings}

As we want to explore the range of low $\tan\beta_1$ and $\tan\beta_2$
we should avoid that the Yukawa couplings become non-perturbative.
The Higgs-Fermion couplings are defined in
Eq.~(\ref{eq:couplingNeutralFerm}) and given for the various types in
Section~\ref{sec:Yukawas}. 
We require
\begin{equation}
  \label{eq:1b}
  \frac{Y^2}{4\pi} < 1 \quad \Rightarrow \quad Y< \sqrt{4\pi}
\end{equation}
for $Y_\tau , Y_b $ and $Y_t$.

\subsubsection{$\Delta M_{b,s}$ Constraints}

We see from
Ref.~\cite{Chakraborti:2021bpy} that
the constraints coming from $\Delta M_{b,s}$ tend to
exclude very low values on $\tan\beta$. Thus, we take
\begin{equation}
  \label{eq:2b}
  \log_{10}(\tan\beta_{1,2}) >-0.5\quad \Rightarrow \quad
  \tan\beta_{1,2} > 10^{-0.5} = 0.31623\, .
\end{equation}

\subsubsection{Limits BR($B\to X_s \gamma$)}

This is a very important bound, for models with charged Higgs bosons.
We follow the discussion of Ref.~\cite{Florentino:2021ybj,Boto:2021qgu},
and following \cite{Akeroyd:2020nfj}, 
we consider 99\% CL (3$\sigma$) for the experimental error:
\begin{equation}
  \label{eq:17}
  2.87 \times 10^{-4} < \text{BR}(B\to X_s \gamma) < 3.77 \times 10^{-4}\, .
\end{equation}

\subsubsection{LHC Constraints}

For the 125GeV scalar, the coupling  modifiers,
are  calculated  directly  from  the  random  angles 
generated and constrained to be within $2\sigma$ of the most
recent ATLAS fit results, \cite[Table 10]{Aad:2019mbh}. 
Having chosen a specific production and decay channel,
the collider event rates can be conveniently  described by the cross section ratios $\mu_{if}^h$,
\beq
\mu_{if}^h=\left(\frac{\sigma_i^{\text{3HDM}}(pp\to h) }{\sigma_i^{\text{SM}}(pp\to h)}\right)\left(\frac{\text{BR}^{\text{3HDM}}(h\to f)}{\text{BR}^{\text{SM}}(h\to f)}\right).
\eeq
Starting from the collision of two protons, the relevant production
mechanisms include: gluon fusion (ggH), vector boson fusion (VBF),
associated production with a vector boson (VH, V = W or Z), and
associated production with a pair of top quarks (ttH). The SM cross
section for the gluon fusion process is calculated using HIGLU
\cite{Spira:1995mt}, and for the other production mechanisms we use
the results of Ref.~\cite{deFlorian:2016spz}. The details can be found
in Ref.\cite{Boto:2021qgu}.

For the heavier neutral and charged scalars,
we use \texttt{HiggsBounds-5.9.1} in Ref.~\cite{Bechtle:2020pkv},
where a list of all the relevant experimental analyses
can be found.
We allow for decays with off-shell scalar bosons,
using the method explained in \cite{Romao:1998sr}.

\section{Results}
\label{sec:results}

\subsection{Comparison of the different BFB conditions}
For each of the symmetries - $U(1) \times U(1)$, $U(1) \times\Z2$,
and $\Z2\times\Z2$ - 
we have generated a large set of points which are consistent
with Al-1 in \eqref{Al-1},
and which pass all current constraints from
$B$-physics, measurements of the 125GeV Higgs properties,
and searches for extra scalars.
We repeated the process for Al-2 in \eqref{Al-2}.

Within each group, we denote by different colours those points
which pass different BFB conditions,
using the following notation:
\begin{itemize}
\item{BFB1} (red points): those points which pass BFB-n but do not pass
BFB-c;
\item{BFB2} (green points): those points which pass BFB-n and also
pass the necessary and sufficient conditions for BFB-c.
These conditions are only known for $U(1)\times U(1)$,
shown in subsection~\ref{subsec:U1U1_BFB_NS},
and for $U(1)\times \Z2$,
shown in subsection~\ref{subsec:U1Z2_BFB_NS}.
The necessary and sufficient conditions for BFB-c are unknown
in the case of $\Z2\times\Z2$ and, thus,
there will be no green points in the corresponding plots.
\item{BFB3} (orange points): those points which pass BFB-c and also
pass the sufficient conditions for BFB-c derived in this article,
but do not pass BFB4 below.
\item{BFB4} (blue points): those points which pass BFB-c and also
pass the sufficient conditions for BFB-c adapted from
those of the $\Z2\times\Z2$ case presented in Ref.~\cite{Grzadkowski:2009bt},
but do not pass BFB3 above.
\item{BFB3+4} (grey points): those points which pass BFB-c and also
pass the sufficient conditions for BFB-c derived in this article,
and in addition also
pass the sufficient conditions for BFB-c adapted from
those of the $\Z2\times\Z2$ case presented in Ref.~\cite{Grzadkowski:2009bt}.
Grey points are the would-be overlap between orange and blue points.
\end{itemize}

%

\begin{figure}[htb] 
  \centering
  \begin{tabular}{cc}
\includegraphics[width=0.47\textwidth]{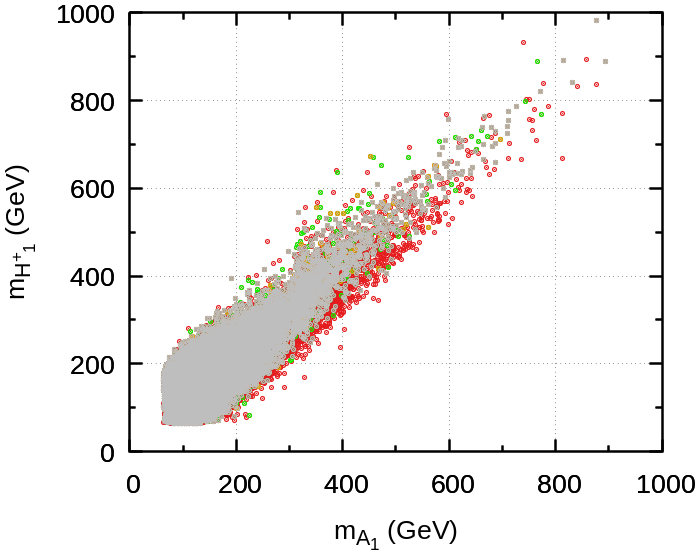}
    &
\includegraphics[width=0.47\textwidth]{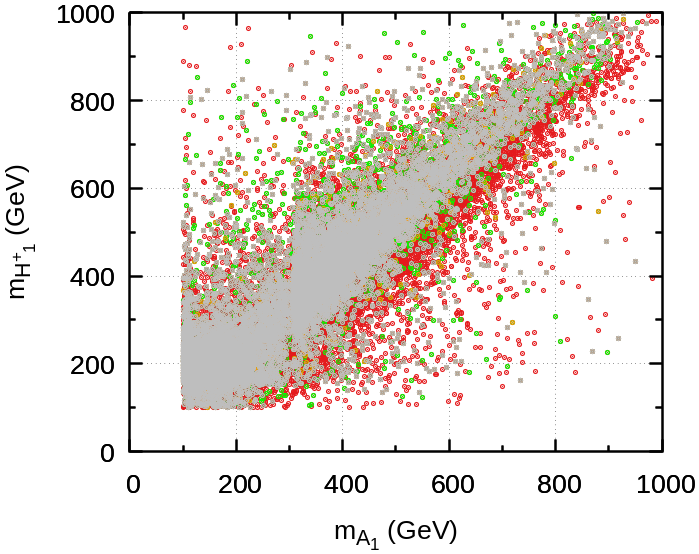}
  \end{tabular}
   \caption{$U(1)\times U(1)$: $(m_{A_1}, m_{H_1^\pm})$ for Type-1 with the
tight Al-2 conditions (left panel) and loose Al-1 conditions (right panel).
BFB1=red, BFB2=green, BFB3=orange, BFB4=blue, BFB3+4=gray}
    \label{fig:ma1-mc1_t1_U1U1}
\end{figure}

We first comment on the difference between the two alignment conditions:
Al-1 and Al-2.
In Figure~\ref{fig:ma1-mc1_t1_U1U1}a,
we show in the $m_{A_1} - m_{H_1^\pm}$ plane
the points which have survived all the constraints and which
have been generated by a 50\% range around the limit~\eqref{eq:44},
as in Eq.~\eqref{Al-2}.
Figure~\ref{fig:ma1-mc1_t1_U1U1}b repeats the exercise
for the much looser alignment constraints in \eqref{Al-1}.
Naturally, points which obey Al-1 but do not obey Al-2 are much
more difficult to generate than points which obey Al-2.
However, as Figure~\ref{fig:ma1-mc1_t1_U1U1}b illustrates,
such points are allowed and correspond to physically interesting
regions of parameter space.
Namely,
and contrary to popular belief,
the oblique parameters do not require degeneracy
within each scalar family.
This confirms and extends results mentioned
in Ref.~\cite{Hernandez-Sanchez:2020aop}
for the case of a very specific DM implementation of $\Z2\times\Z2$.

\begin{figure}[htb]
  \centering
  \begin{tabular}{cc}
\includegraphics[width=0.47\textwidth]{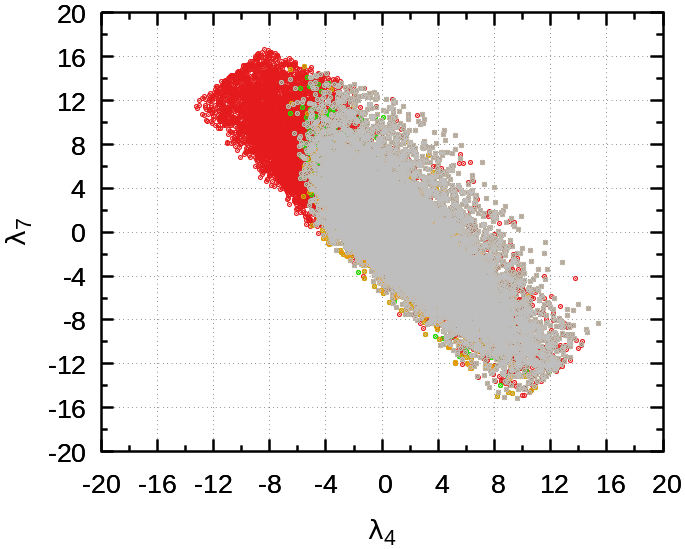}
    &
\includegraphics[width=0.47\textwidth]{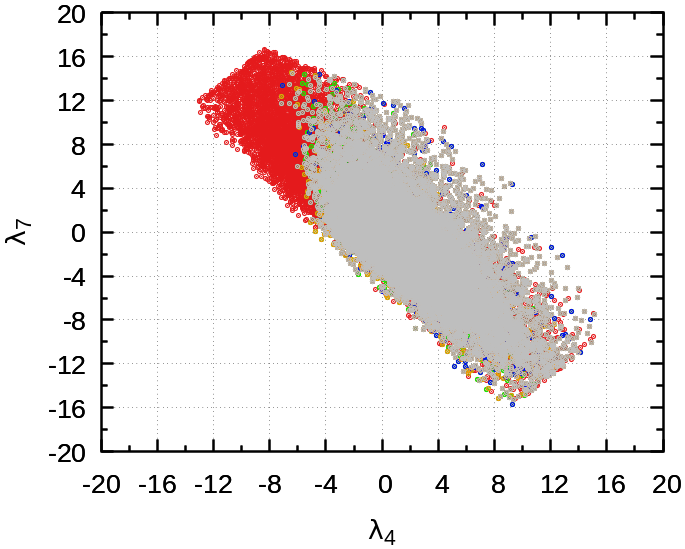}
  \end{tabular}
   \caption{Left panel:    $U(1)\times U(1)$: $(\lambda_4, \lambda_7)$ for Type-1 with the Al-2 conditions.\\ Right panel: $U(1)\times \Z2$: $(\lambda_4, \lambda_7)$ for Type-1 with the Al-2 conditions.}
    \label{fig:L4-L7_t1_U1U1_U1Z2_k16}
\end{figure}

\begin{figure}[htb]
\centering
\includegraphics[width = 0.5\textwidth]{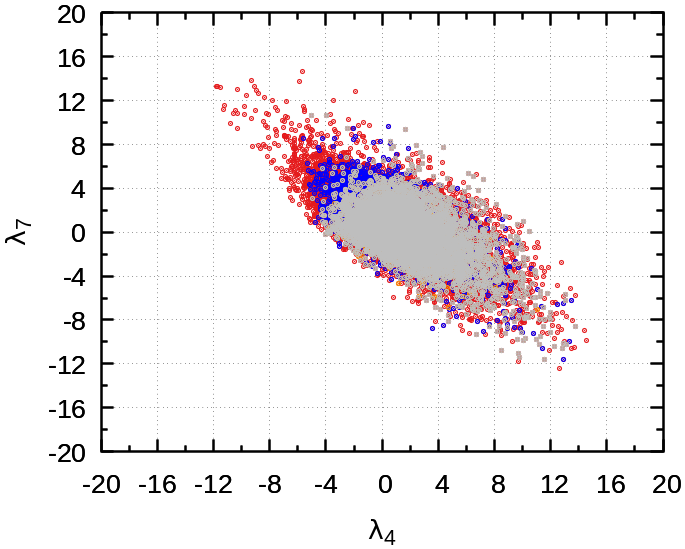}
\caption{$\Z2\times \Z2$: $(\lambda_4, \lambda_7)$ for Type-1 with the
Al-2 conditions.}
    \label{fig:L4-L7_t1_Z2Z2_k16}
\end{figure}

Now we turn to a second important issue.
Could it be that by using only BFB-n, without concern
about BFB-c one is led into wrong physical conclusions?
After all, it could be that points which are BFB-n
but not BFB-c do not differ in their physical
consequences from points which obey both BFB-n
and BFB-c.
This is not the case, as we illustrate in 
Figures~\ref{fig:L4-L7_t1_U1U1_U1Z2_k16}a,
\ref{fig:L4-L7_t1_U1U1_U1Z2_k16}b,
and \ref{fig:L4-L7_t1_Z2Z2_k16}.
We see that points which are BFB-n but not BFB-c
(BFB1-red points) do allow for a negative and large $\lambda_4$,
together with a positive and large $\lambda_7$.
The same type of features appear in the
$\lambda_5-\lambda_8$ plane.
And this occurs for all symmetries studied in this article:
$U(1)\times U(1)$ in Figure~\ref{fig:L4-L7_t1_U1U1_U1Z2_k16}a;
$U(1)\times \Z2$ in Figure~\ref{fig:L4-L7_t1_U1U1_U1Z2_k16}b;
and $\Z2\times \Z2$ in Figure~\ref{fig:L4-L7_t1_Z2Z2_k16}.
We thus conclude that ignoring BFB-c does lead
to wrong physical conclusions.
Dealing with the charge breaking directions is not an option;
it is a must.

A third and curious conclusion arises from the implementation
of the different BFB constraints.
In the $U(1)\times U(1)$ and $U(1)\times \Z2$ cases there
are three BFB conditions of interest.
The true necessary and sufficient BFB conditions
in subsection 
\ref{subsec:U1U1_BFB_NS},
and \ref{subsec:U1Z2_BFB_NS},
respectively;
the conditions proposed in this article;
and the adaptation of the
sufficient conditions for BFB-c presented for the $\Z2\times\Z2$
case in Ref.~\cite{Grzadkowski:2009bt}.

We start by noticing that the green points in
Figures~\ref{fig:ma1-mc1_t1_U1U1}a,
\ref{fig:ma1-mc1_t1_U1U1}b,
\ref{fig:L4-L7_t1_U1U1_U1Z2_k16}a,
and \ref{fig:L4-L7_t1_U1U1_U1Z2_k16}b,
do not seem to occupy regions of parameter space
far different from those allowed by the more stringent
sufficient conditions BFB3, BFB4, and BFB3+4.
This is a first hint that maybe using sufficient conditions
does not skew the physical interpretation of the models.
We will come back to this issue below.
Interestingly,
there seem to be no single blue point in the
Figures \ref{fig:ma1-mc1_t1_U1U1}a,
\ref{fig:ma1-mc1_t1_U1U1}b,
and
\ref{fig:L4-L7_t1_U1U1_U1Z2_k16}a,
corresponding to the $U(1) \times U(1)$ case.
Indeed, we have found numerically that all
points which obey the adaptation to
$U(1) \times U(1)$ of the 
sufficient BFB-c conditions presented for the $\Z2\times\Z2$
case in Ref.~\cite{Grzadkowski:2009bt}
also obey the $U(1)\times U(1)$ sufficient BFB-c
conditions proposed by us in subsection~\ref{subsec:U1U1_BFB_us}.
This is illustrated by the gray points.
The converse is not true.
Thus we find in Figures \ref{fig:ma1-mc1_t1_U1U1}a,
\ref{fig:ma1-mc1_t1_U1U1}b,
and
\ref{fig:L4-L7_t1_U1U1_U1Z2_k16}a orange points,
which correspond to points which pass the sufficient BFB-c
conditions of subsection~\ref{subsec:U1U1_BFB_us} but do
\textit{not} pass the sufficient BFB-c
conditions of subsection~\ref{subsec:U1U1_BFB_Grz}.
In contrast,
Figure~\ref{fig:L4-L7_t1_U1U1_U1Z2_k16}b,
which correspond to the $U(1)\times \Z2$ case,
contain: i) points in gray which pass both sets of bounds;
ii) points in orange which pass the conditions of 
subsection~\ref{subsec:U1U1_BFB_us} but do
\textit{not} pass the conditions of subsection~\ref{subsec:U1U1_BFB_Grz};
but also iii) points in blue which pass the conditions of 
subsection~\ref{subsec:U1U1_BFB_Grz} but do
\textit{not} pass the conditions of subsection~\ref{subsec:U1U1_BFB_us}.
In fact,
we find the quite curious result that our simulation
consistently generated more blue points than orange points.
This is even more apparent in
Figure~\ref{fig:L4-L7_t1_Z2Z2_k16} concerning the $\Z2\times\Z2$ case.
That figure is based on one simulation where we found 
6977 BFB3 orange points,
10087 BFB4 blue points,
and 6842 BFB3+4 gray overlap points.
This could be due to the following.
The sufficient conditions for the $\Z2\times\Z2$ case were found in
Ref.~\cite{Grzadkowski:2009bt} by a careful study of the
charge breaking directions of that specific potential.
Thus, it is not surprising that they are more helpful in that case
than in the $U(1)\times U(1)$ or $U(1)\times \Z2$ cases.

\begin{figure}[htb]
  \centering
  \begin{tabular}{cc}
\includegraphics[width=0.47\textwidth]{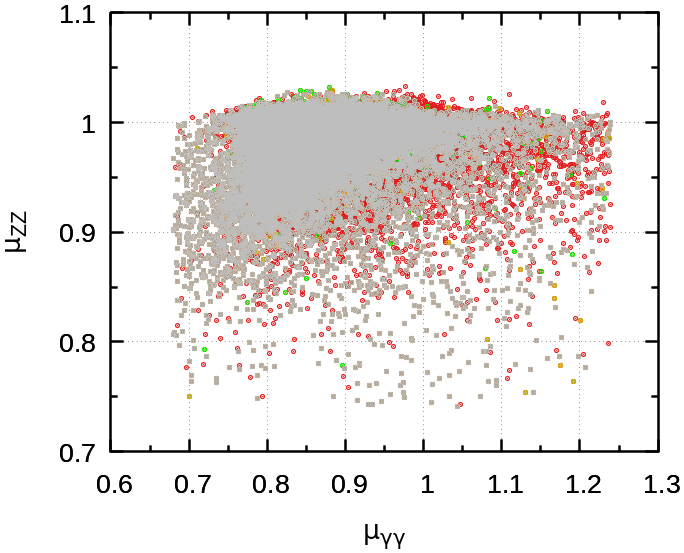}
    &
\includegraphics[width=0.47\textwidth]{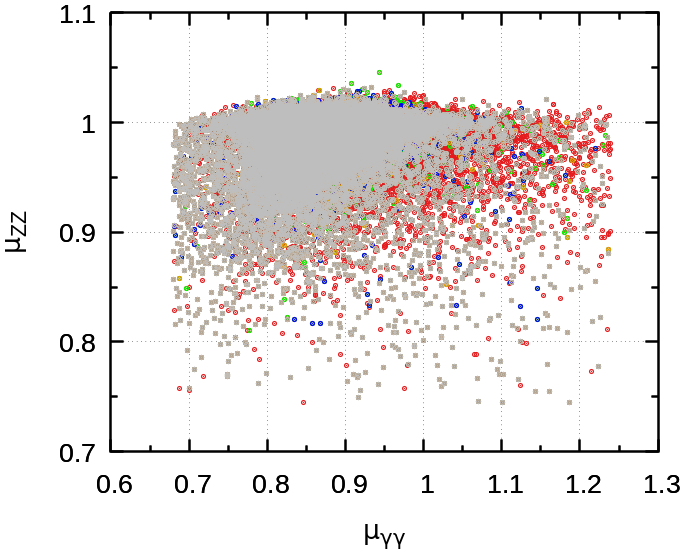}
  \end{tabular}
   \caption{Left panel:    $U(1) \times U(1)$: $\mu_{ZZ}-\mu_{\gamma\gamma}$
    plane for the gluon fusion production channel.
    For Type-1 with Al-2 conditions.\\ Right panel: $U(1) \times\Z2$: $\mu_{ZZ}-\mu_{\gamma\gamma}$
    plane for the gluon fusion production channel.
    For Type-1 with Al-2 conditions.}
    \label{fig:gaga-zz_t1_U1U1_U1Z2_k16}
\end{figure}

\begin{figure}[htb]
\centering
\includegraphics[width = 0.5\textwidth]{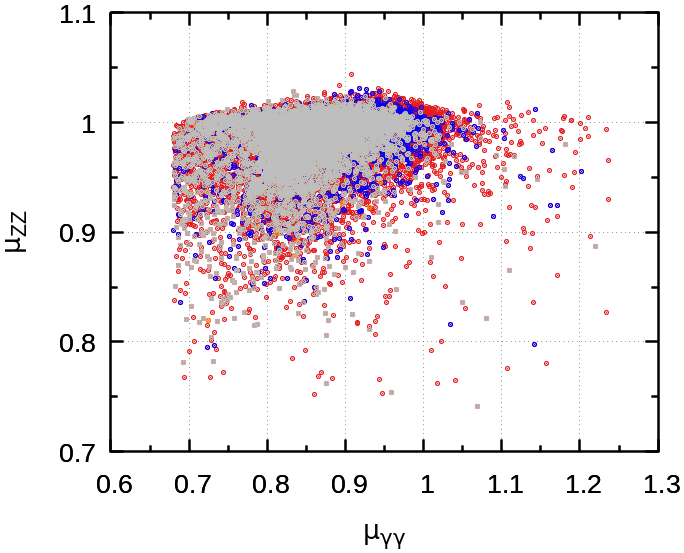}
\caption{$\Z2 \times \Z2$: $\mu_{ZZ}-\mu_{\gamma\gamma}$
    plane for the gluon fusion production channel.
    For Type-1 with Al-2 conditions.}
    \label{fig:gaga-zz_t1_Z2Z2_k16}
\end{figure}

\begin{figure}[htb]
  \centering
  \begin{tabular}{cc}
\includegraphics[width=0.47\textwidth]{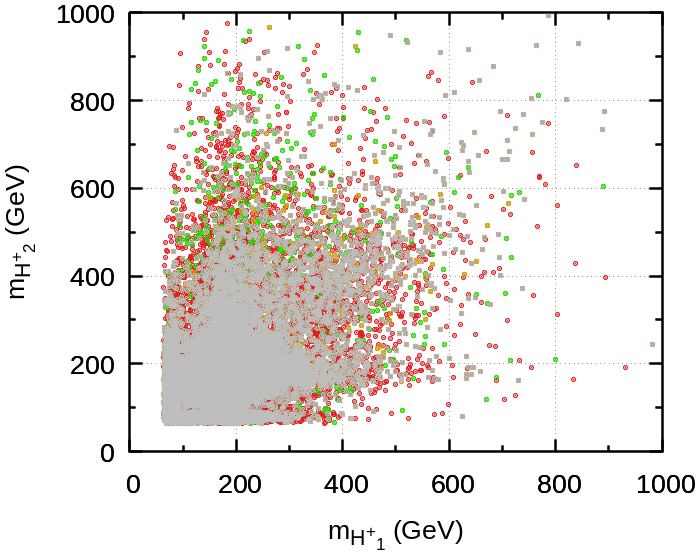}
    &
\includegraphics[width=0.47\textwidth]{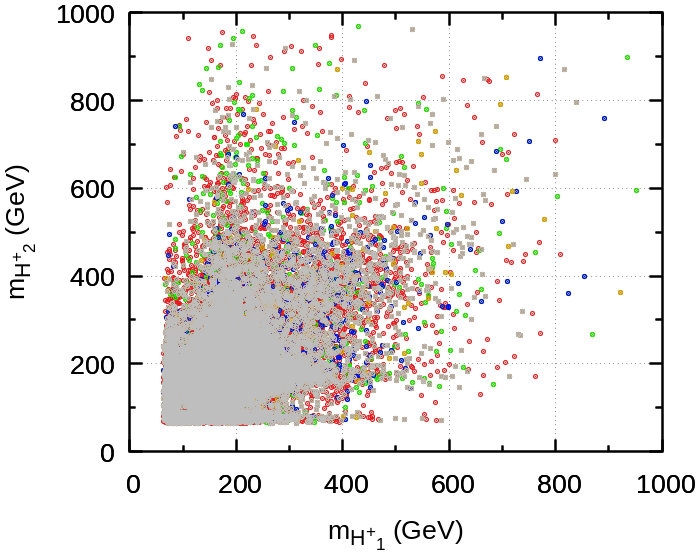}
  \end{tabular}
   \caption{Left panel:    $U(1) \times U(1)$: charged scalar masses for Type-1 with Al-2 conditions.\\ Right panel: $U(1) \times \Z2$: charged scalar masses for Type-1 with Al-2 conditions.}
    \label{fig:mc1-mc2_t1_U1U1_U1Z2_k16}
\end{figure}

\begin{figure}[htb]
\centering
\includegraphics[width = 0.5\textwidth]{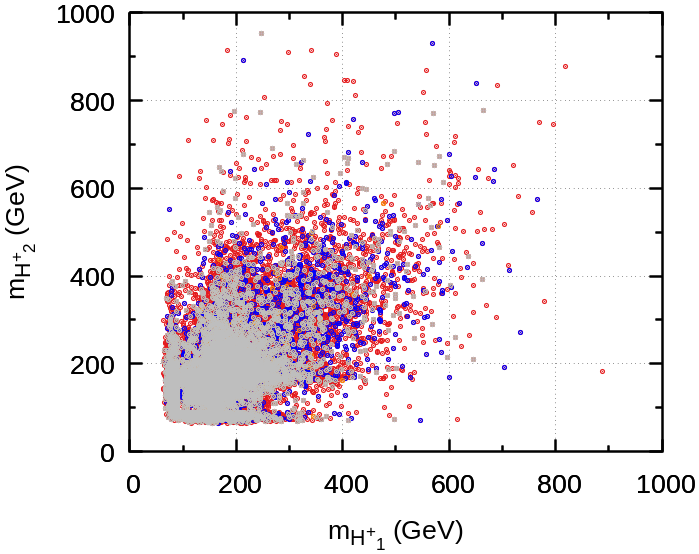}
\caption{$\Z2 \times \Z2$: charged scalar masses for Type-1 with Al-2 conditions.}
    \label{fig:mc1-mc2_t1_Z2Z2_k16}
\end{figure}

We now turn to the question of whether using sufficient BFB-c
instead of the correct necessary and sufficient BFB-c conditions
does (or not) constrain unduly the physical quantities.
We have calculated all consequences of the various points found
for the 125GeV scalar and for searches into heavier scalars.
Next we plotted all pairs of observables in the respective
planes, looking for physical differences between the placement
of the green points versus points with sufficient BFB-c conditions.
We have found no evidence of a difference.
We illustrate such searches below.
We show the $\mu_{\gamma\gamma} - \mu_{ZZ}$ plane
in Figures~\ref{fig:gaga-zz_t1_U1U1_U1Z2_k16}a,
\ref{fig:gaga-zz_t1_U1U1_U1Z2_k16}b,
and \ref{fig:gaga-zz_t1_Z2Z2_k16},
for the $U(1)\times U(1)$,
$U(1)\times \Z2$,
and $\Z2\times \Z2$ symmetries, respectively.
The exercise is repeated for the $m_{H_1^\pm} - m_{H_2^\pm}$
plane in Figures \ref{fig:mc1-mc2_t1_U1U1_U1Z2_k16}a,
\ref{fig:mc1-mc2_t1_U1U1_U1Z2_k16}b,
and \ref{fig:mc1-mc2_t1_Z2Z2_k16}.

There is no physical difference that can be considered statistically
significant; there are only minor differences between placement of colours,
due to the sparse placement of a (necessarily) limited numerical
simulation\footnote{See a more detailed discussion of this point
in the next section.}.
By looking at hundreds of such plots we conclude that:
\begin{enumerate}
\item Using BFB-n bounds while ignoring BFB-c considerations does lead
to wrong physical conclusions.
\item In contrast, using safe sufficient BFB-c bounds versus using
(when available) the exact necessary and sufficient BFB-c conditions
does not seem to introduce a bias in the physical observables.
\item Moreover, using different safe BFB-c bounds does affect the
number of points generated (for equal running time) but it does not
seem to introduce a bias in the analysis.
\end{enumerate}

\subsection{Details of the numerical simulation}
When discussing
Figs.~\ref{fig:mc1-mc2_t1_U1U1_U1Z2_k16}--\ref{fig:mc1-mc2_t1_Z2Z2_k16}
we mentioned that the high mass region was sparsely populated. In those
figures we were using the conditions of Eq.~(\ref{Al-2}). As these also
include the exact conditions of
Eq.~(\ref{eq:44}), one would not expect a difficulty
in having high masses. To explain this point one must realize first,
that the points in those figures have passed all the cuts imposed by the
theoretical and experimental constraints. Secondly, and more
important, when we randomly sample as in  Eq.~(\ref{Al-2}), it is very
unlikely that we get any point that is close to the conditions of
Eq.~(\ref{eq:44}). 

To better illustrate this point, we now focus on the $\Z2 \times \Z2$ model
with the lowest bound developed in this article.
We generated four not overlapping
sets, where instead of Eq.~(\ref{Al-2}), we use the following intervals,
\begin{itemize}
\item $\text{Set A}: [0\%,1\%]\ \text{red}$
\item $\text{Set B}: [1\%,10\%]\ \text{green}$
\item $ \text{Set C}: [10\%,20\%]\ \text{orange}$
\item $\text{Set D}: [20\%,50\%]\  \text{blue}$
\end{itemize}

\begin{figure}[htb]  
  \centering
  \begin{tabular}{cc}
\includegraphics[width=0.47\textwidth]{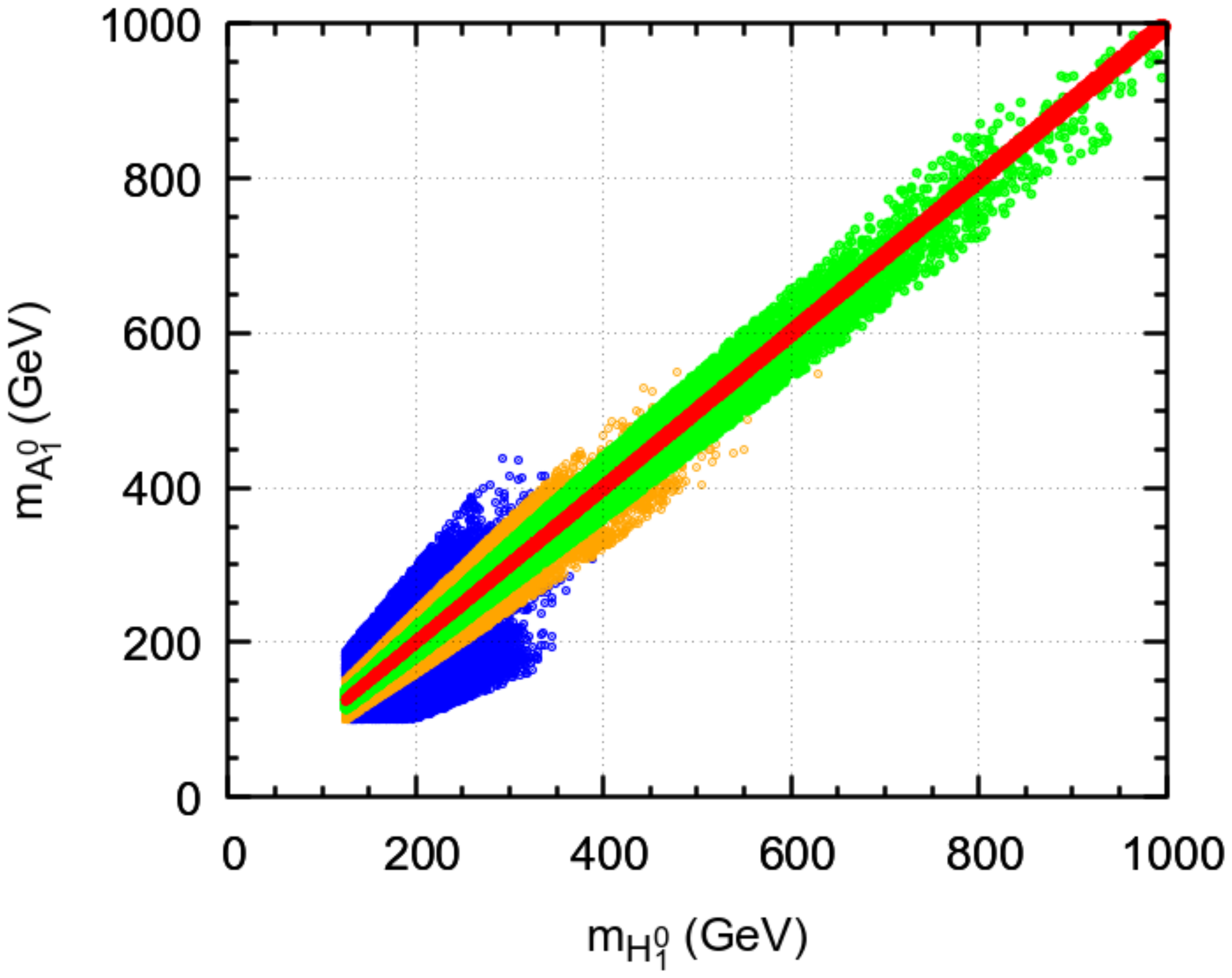}
    &
\includegraphics[width=0.47\textwidth]{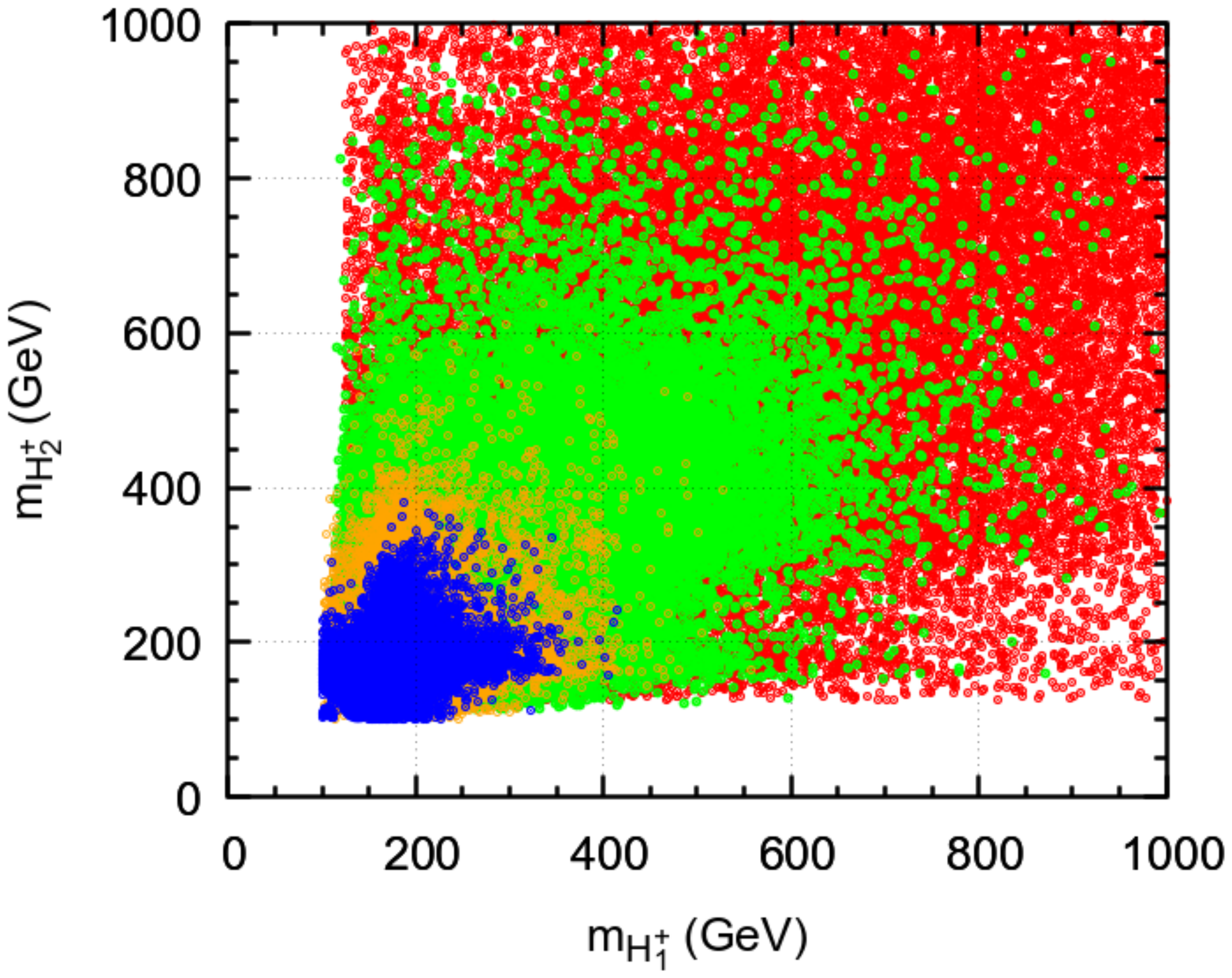}      
  \end{tabular}
  \caption{Left panel: $m_{H_{1}}$ versus $m_{A_{1}}$ for the four
  sets indicated in the text.\\ Right panel: Same for $m_{H^+_{1}}$
  versus $m_{H^+_{1}}$.}
\label{fig:masses}
\end{figure}

In the left panel of Fig.~\ref{fig:masses} we plot the mass of $A_1$
versus the mass of $H_1$. In the conditions of Eq.~(\ref{eq:44}), this
would just be a straight line. Here, for Set A, we are very close to
those conditions. As we move away from  Eq.~(\ref{eq:44}), we notice
two things. First, the line moves into a broader band. Second, larger
masses are being cut. This is especially true for Set~D, where we are
more than 20\% away from Eq.~(\ref{eq:44}). This is due to the fact
that we are now far away from the symmetric conditions of
Eq.~(\ref{eq:47}), and therefore from the quasi-SM situation.
Then, the
combination of constraints, including the LHC results, makes it
increasingly difficult to generate good points with large masses. We
show this in a different way in the right panel of
Fig.~\ref{fig:masses}. We see that high masses are easy to be
generated for smaller deviations from the symmetric situation of
Eq.~(\ref{eq:44}). We notice that there is no contradiction between the
right panel of Fig.~\ref{fig:masses} and
Fig.~\ref{fig:mc1-mc2_t1_Z2Z2_k16}. In fact the points for this figure
were generated according to Eq.~(\ref{Al-2}), which means that this
includes both points close to the symmetric limit of 
Eq.~(\ref{eq:44}) and points deviating from it up to 50\%.

The reason for including points away from the symmetric limit is that
otherwise we get results that are very close to the SM, with very
little room for new phenomenology. 

\begin{figure}[htb]
  \centering
  \begin{tabular}{cc}
\includegraphics[width=0.47\textwidth]{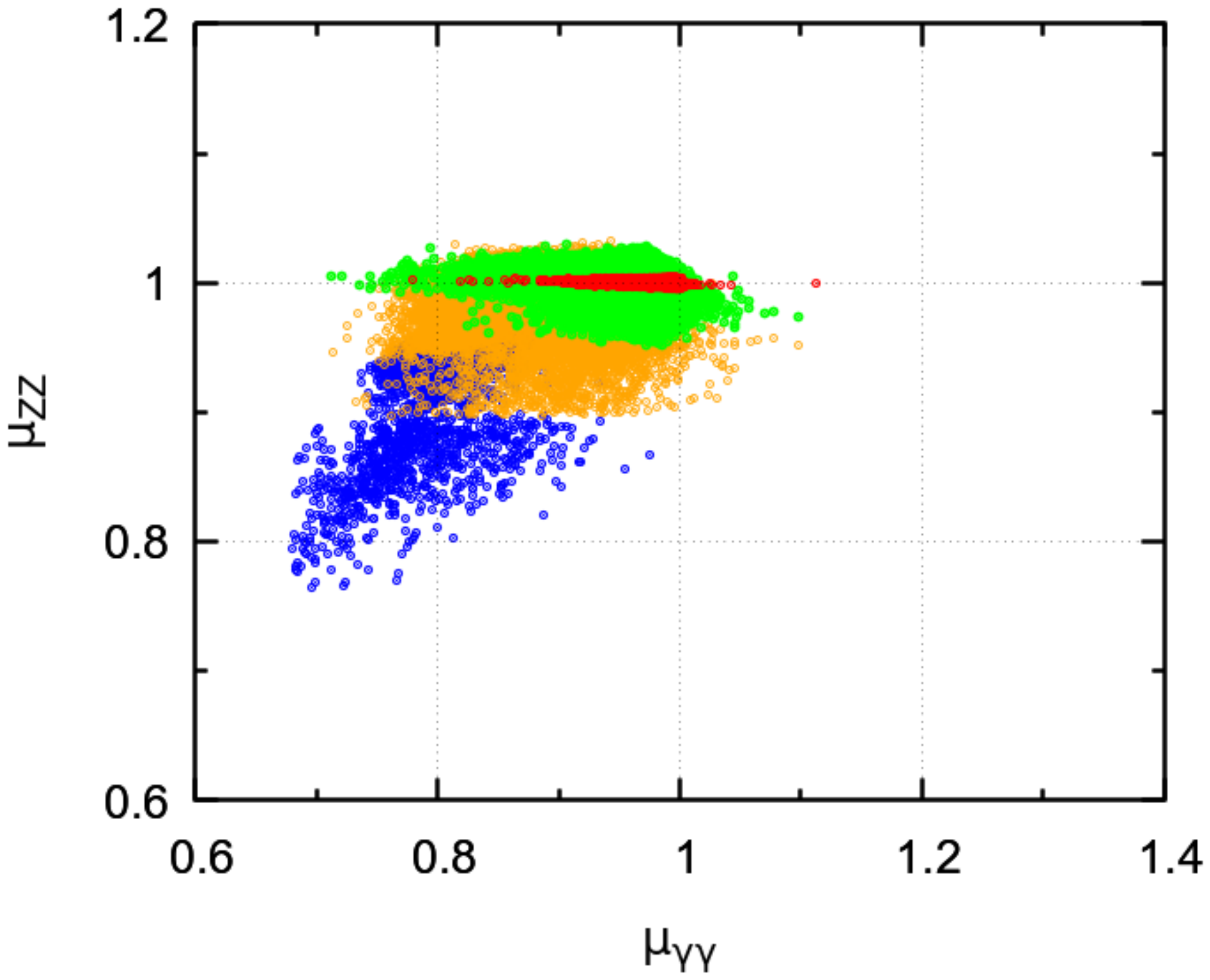}
    &
\includegraphics[width=0.47\textwidth]{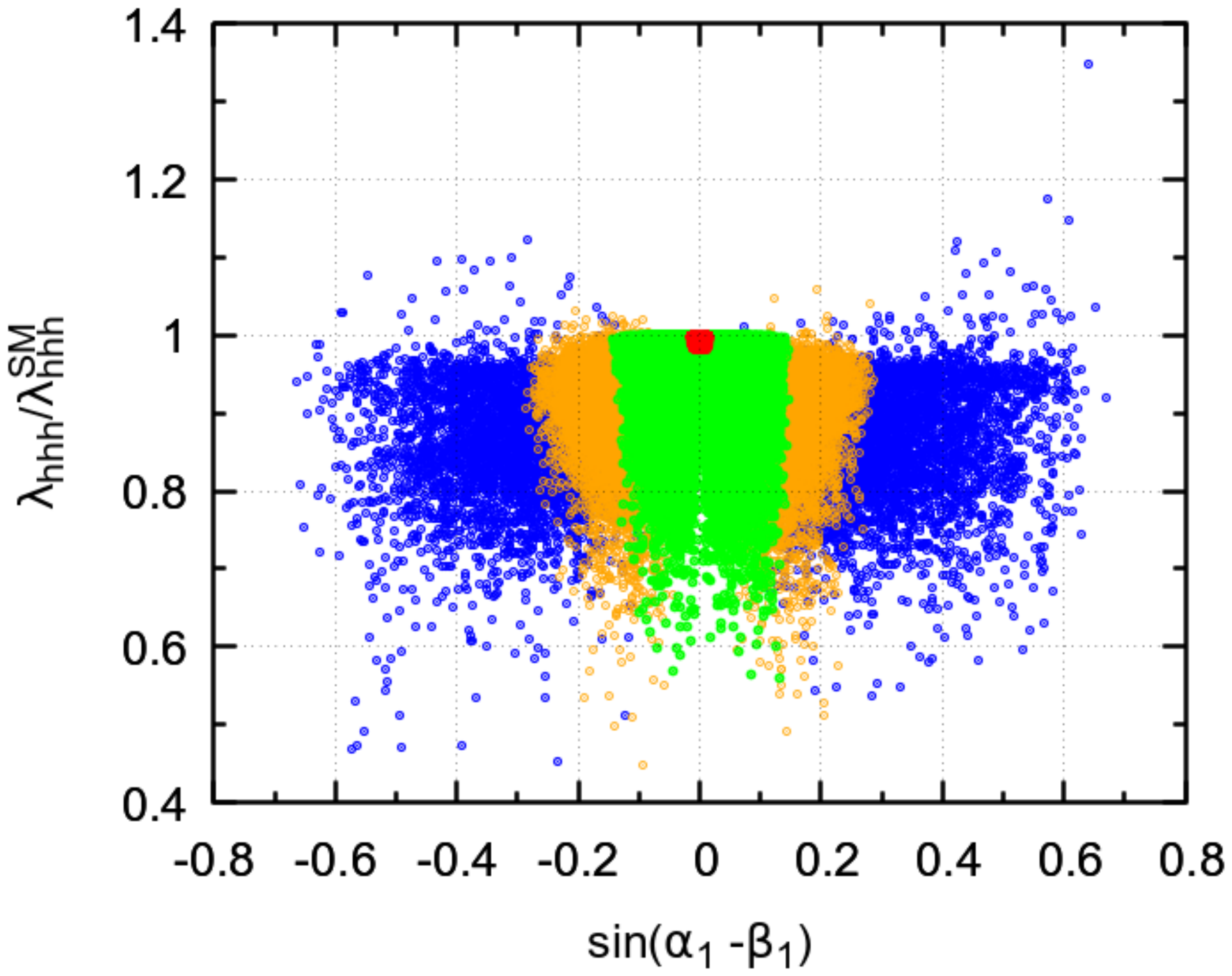}      
  \end{tabular}
  \caption{Left panel: $\mu_{\gamma\gamma}$ versus $\mu_{ZZ}$ for the four
  sets indicated in the text.\\ Right panel: Same for
  $\lambda_{hhh}/\lambda_{hhh}^{\rm SM}$
  versus $\sin(\alpha_1-\beta_1)$.}
  \label{fig:observables}
\end{figure}

This is illustrated in Fig.~\ref{fig:observables} where in the left
panel we plot the signal strengths $\mu_{ZZ}$ versus
$\mu_{\gamma\gamma}$ for the same four sets. In the right panel we
plot the ratio of the $\lambda_{hhh}$ coupling to the SM. In both
cases we see that, if we remain too close to the symmetric limit of
Eq.~(\ref{eq:44}), the results are very close to the SM, especially for
the Higgs boson triple coupling.

One final comment is in order. When we compare the left panel of
Fig.~\ref{fig:observables} with Fig.~\ref{fig:gaga-zz_t1_Z2Z2_k16}, we
see that the former is included in the later, but covers a smaller
region. The reason again is that our four sets are more constrained
than the generation in Eq.~(\ref{Al-2}). For instance one can have
some of the parameters at 50\% and others at 10\%, all obeying 
Eq.~(\ref{Al-2}).
But, in the logic of the four sets used in this section, they would be
included in none. So, the final lesson is that we should try to be as
far away as possible from the symmetric limit (but still compatible
with all constraints) in order to have a richer beyond the SM phenomenology.
And, that such large deviations are still compatible with all
present theoretical and experimental bounds.

\section{\label{sec:conclusions}Conclusions}

Most models of physics beyond the SM include an extended scalar sector.
The phenomenology of such models cannot be reliably analyzed before a
careful assessment of whether the potential is bounded from below
and whether the chosen solution of the stationarity equations
is indeed a global minimum. In this article, we address the first issue,
stressing the necessity of studying both the neutral and the charge
breaking directions.

First, we develop a strategy to find sufficient BFB conditions.
It hinges on finding a potential which lies below the original potential
and to which the positivity conditions
\cite{Klimenko:1984qx,Kannike:2012pe} can be applied.
We study in detail the $U(1) \times U(1)$, $U(1) \times \Z2$,
and $\Z2\times\Z2$ 3HDMs, for both BFB-n and BFB-c.
Then, we adapt the sufficient BFB $\Z2\times\Z2$ results of
\cite{Grzadkowski:2009bt} to the $U(1)\times U(1)$
and $U(1)\times \Z2$ 3HDM,
comparing with our bounds and highlighting both the similarities
(gray points in Figs.~\ref{fig:ma1-mc1_t1_U1U1}
through \ref{fig:mc1-mc2_t1_Z2Z2_k16})
and the differences
(orange and blue points in Figs.~\ref{fig:ma1-mc1_t1_U1U1}
through \ref{fig:mc1-mc2_t1_Z2Z2_k16}).

Second,
we address the impact that the choice of sufficient (but not necessary)
BFB conditions might have on phenomenological studies.
It could be that the sufficient conditions used in these models
exclude good points that would yield dramatically new features.
This study can be performed in the $U(1) \times U(1)$
and $U(1) \times \Z2$ cases, where the correct necessary and
sufficient BFB conditions are possible \cite{Faro:2019vcd,Faro:2019}.
This required us to set up the full model, including the Yukawa
couplings, which we take to be consistent with the absence of
flavour changing neutral scalar couplings
\cite{Ferreira:2010xe,Yagyu:2016whx}.
We present in detail all couplings,
but concentrate our phenomenological studies
on Type-I models.
This facilitates the scrutiny of our results and
also facilitates further detailed studies of specific aspects
of the phenomenology of these models, for all Types of Yukawa couplings.

After analyzing hundreds of correlations in two-dimensional planes
of experimental observables, we find no evidence that points
allowed by the complete necessary and sufficient BFB conditions but
excluded by our sufficient BFB bounds would yield any new phenomenological
features.
We did this for both $U(1) \times U(1)$ and $U(1) \times \Z2$.
A few examples are shown in Sec.~\ref{sec:results}.
Although not an airtight proof,
as is the case in any numerical simulation,
our results provide some reassurance that the sufficient BFB conditions
developed here do not significantly skew the phenomenology in cases
where no complete necessary and sufficient conditions are known,
such as the $\Z2 \times \Z2$ and $\Z3$ 3HMDs.

\vspace{1ex}

\section*{Acknowledgments}
\noindent
We are grateful to Dipankar Das for useful suggestions.
This work is supported in part by the Portuguese
Funda\c{c}\~{a}o para a Ci\^{e}ncia
e Tecnologia\/ (FCT) under Contracts
CERN/FIS-PAR/0008/2019,
PTDC/FIS-PAR/29436/2017,
UIDB/00777/2020,
and UIDP/00777/2020;
these projects are partially funded through POCTI (FEDER),
COMPETE,
QREN,
and the EU.
The work of R. Boto is also supported
by FCT with the PhD grant PRT/BD/152268/2021.

\appendix
\section{\label{app:lambdas}Potential parameters in terms of physical variables}

We list here the relation of the parameters of the potential and
masses and angles for the three cases.

\subsection{The $U(1)\times U(1)$ potential}

As there are no $\lambda''_{10}$, $\lambda''_{11}$,  and $\lambda''_{12}$,
we can also solve for the three  soft terms.  The expressions are
\begin{align} 
\label{eq:41} 
  \lambda_1=& 
\frac{1}{2 c_{\beta_{1}}^3
  c_{\beta_{2}}^3 v^2}
\left[c_{\alpha_{1}}^2 c_{\alpha_{2}}^2 c_{\beta_{1}} c_{\beta_{2}}
m_{h}^2+c_{\alpha_{1}}^2 c_{\alpha_{3}}^2
   c_{\beta_{1}} c_{\beta_{2}} m_{H_{2}}^2 s_{\alpha_{2}}^2+c_{\alpha_{1}}^2
   c_{\beta_{1}} c_{\beta_{2}} m_{H_{1}}^2 s_{\alpha_{2}}^2
   s_{\alpha_{3}}^2\right.\nonumber\\
&\left.
   +2
   c_{\alpha_{1}} c_{\alpha_{3}} c_{\beta_{1}} c_{\beta_{2}} m_{H_{1}}^2 s_{\alpha_{1}}
   s_{\alpha_{2}} s_{\alpha_{3}}-2 c_{\alpha_{1}} c_{\alpha_{3}} c_{\beta_{1}} c_{\beta_{2}}
   m_{H_{2}}^2 s_{\alpha_{1}} s_{\alpha_{2}} s_{\alpha_{3}}+c_{\alpha_{3}}^2
   c_{\beta_{1}} c_{\beta_{2}} m_{H_{1}}^2 s_{\alpha_{1}}^2\right.\nonumber\\
&\left.
   +c_{\beta_{1}}
   c_{\beta_{2}} m_{H_{2}}^2 s_{\alpha_{1}}^2 s_{\alpha_{3}}^2+c_{\beta_{2}}
   m_{12}^2 s_{\beta_{1}}+m_{13}^2 s_{\beta_{2}}\right]\, ,
  \\[+2mm]
  \lambda_2=&\frac{1}{2
    c_{\beta_{2}}^3 s_{\beta_{1}}^3 v^2}
  \left[c_{\alpha_{1}}^2 c_{\alpha_{3}}^2 c_{\beta_{2}} m_{H_{1}}^2
  s_{\beta_{1}}+c_{\alpha_{1}}^2 c_{\beta_{2}}
   m_{H_{2}}^2 s_{\alpha_{3}}^2 s_{\beta_{1}}-2 c_{\alpha_{1}} c_{\alpha_{3}}
   c_{\beta_{2}} m_{H_{1}}^2 s_{\alpha_{1}} s_{\alpha_{2}} s_{\alpha_{3}}
   s_{\beta_{1}}\right.\nonumber\\
&\left.\hskip 15mm
   +2 c_{\alpha_{1}} c_{\alpha_{3}} c_{\beta_{2}} m_{H_{2}}^2
   s_{\alpha_{1}} s_{\alpha_{2}} s_{\alpha_{3}} s_{\beta_{1}}
   +c_{\alpha_{2}}^2 c_{\beta_{2}} m_{h}^2 s_{\alpha_{1}}^2
   s_{\beta_{1}}+c_{\alpha_{3}}^2 c_{\beta_{2}} 
   m_{H_{2}}^2 s_{\alpha_{1}}^2 s_{\alpha_{2}}^2 s_{\beta_{1}}
 \right.\nonumber\\
 &\left.\hskip 15mm
   +c_{\beta_{1}}
   c_{\beta_{2}} m_{12}^2+c_{\beta_{2}} m_{H_{1}}^2 s_{\alpha_{1}}^2
   s_{\alpha_{2}}^2 s_{\alpha_{3}}^2 s_{\beta_{1}}+m_{23}^2 s_{\beta_{2}}\right]\, ,
  \\[+2mm]
  \lambda_3=&\frac{1}{2
   s_{\beta_{2}}^3 v^2}\left[c_{\alpha_{2}}^2 c_{\alpha_{3}}^2 m_{H_{2}}^2
  s_{\beta_{2}}+c_{\alpha_{2}}^2 m_{H_{1}}^2 s_{\alpha_{3}}^2
   s_{\beta_{2}}+c_{\beta_{1}} c_{\beta_{2}} m_{13}^2+c_{\beta_{2}}
   m_{23}^2 s_{\beta_{1}}+m_{h}^2 s_{\alpha_{2}}^2 s_{\beta_{2}}\right]
  \\[+2mm]
  \lambda_4=&\frac{1}{c_{\beta_{1}} c_{\beta_{2}}^2 s_{\beta_{1}} v^2}
  \left[-c_{\alpha_{1}}^2 c_{\alpha_{3}} m_{H_{1}}^2 s_{\alpha_{2}}
  s_{\alpha_{3}}+c_{\alpha_{1}}^2 c_{\alpha_{3}}
   m_{H_{2}}^2 s_{\alpha_{2}} s_{\alpha_{3}}+c_{\alpha_{1}} c_{\alpha_{2}}^2
   m_{h}^2 s_{\alpha_{1}}\right.\nonumber\\
&\left.\hskip 20mm
   -c_{\alpha_{1}} c_{\alpha_{3}}^2 m_{H_{1}}^2
   s_{\alpha_{1}}+c_{\alpha_{1}} c_{\alpha_{3}}^2 m_{H_{2}}^2 s_{\alpha_{1}}
   s_{\alpha_{2}}^2+c_{\alpha_{1}} m_{H_{1}}^2 s_{\alpha_{1}} s_{\alpha_{2}}^2
   s_{\alpha_{3}}^2 \right.\nonumber\\
&\left.\hskip 20mm
 -c_{\alpha_{1}} m_{H_{2}}^2 s_{\alpha_{1}}
   s_{\alpha_{3}}^2+c_{\alpha_{3}} m_{H_{1}}^2 s_{\alpha_{1}}^2 s_{\alpha_{2}}
   s_{\alpha_{3}}-c_{\alpha_{3}} m_{H_{2}}^2 s_{\alpha_{1}}^2 s_{\alpha_{2}}
   s_{\alpha_{3}}\right.\nonumber\\
&\left.\hskip 20mm
   -c_{\beta_{1}} c_{\beta_{2}}^2 \lambda_{7} s_{\beta_{1}}
   v^2-m_{12}^2 \right]\, ,
  \\[+2mm]
  \lambda_5=&\frac{1}{c_{\beta_{1}} c_{\beta_{2}} s_{\beta_{2}} v^2}
  \left[-c_{\alpha_{1}} c_{\alpha_{2}} c_{\alpha_{3}}^2 m_{H_{2}}^2
  s_{\alpha_{2}}+c_{\alpha_{1}} c_{\alpha_{2}}
   m_{h}^2 s_{\alpha_{2}}-c_{\alpha_{1}} c_{\alpha_{2}} m_{H_{1}}^2
   s_{\alpha_{2}} s_{\alpha_{3}}^2\right.\nonumber\\
&\left.\hskip 20mm
   -c_{\alpha_{2}} c_{\alpha_{3}} m_{H_{1}}^2
   s_{\alpha_{1}} s_{\alpha_{3}}+c_{\alpha_{2}} c_{\alpha_{3}} m_{H_{2}}^2 s_{\alpha_{1}}
   s_{\alpha_{3}}-c_{\beta_{1}} c_{\beta_{2}} \lambda_{8} s_{\beta_{2}}
   v^2-m_{13}^2\right]\, ,
  \\[+2mm]
  \lambda_6=&\frac{1}{c_{\beta_{2}} s_{\beta_{1}} s_{\beta_{2}} v^2}
  \left[c_{\alpha_{1}} c_{\alpha_{2}} c_{\alpha_{3}} m_{H_{1}}^2
  s_{\alpha_{3}}-c_{\alpha_{1}} c_{\alpha_{2}} c_{\alpha_{3}}
   m_{H_{2}}^2 s_{\alpha_{3}}-c_{\alpha_{2}} c_{\alpha_{3}}^2 m_{H_{2}}^2
   s_{\alpha_{1}} s_{\alpha_{2}}\right.\nonumber\\
&\left.\hskip 20mm
   +c_{\alpha_{2}} m_{h}^2 s_{\alpha_{1}}
   s_{\alpha_{2}}-c_{\alpha_{2}} m_{H_{1}}^2 s_{\alpha_{1}} s_{\alpha_{2}}
   s_{\alpha_{3}}^2-c_{\beta_{2}} \lambda_{9} s_{\beta_{1}} s_{\beta_{2}}
   v^2-m_{23}^2 \right]\, ,
  \\[+2mm]
  \lambda_7=&-\frac{2}{c_{\beta_{1}}
    c_{\beta_{2}}^2 s_{\beta_{1}} v^2 }
  \left[ c_{\beta_{1}}^2 \left(c_{\gamma_{2}} s_{\beta_{2}}
  s_{\gamma_{2}}
   \left(m_{H_2^\pm}^2-m_{H_1^\pm}^2\right)+m_{12}^2\right)+c_{\beta_{1}}
   s_{\beta_{1}} \left(c_{\gamma_{2}}^2 \left(m_{H_1^\pm}^2-m_{H_2^\pm}^2
       s_{\beta_{2}}^2\right)\right.\right.\nonumber\\
&\left.\left.\hskip 20mm
     +s_{\gamma_{2}}^2 \left(m_{H_2^\pm}^2-m_{H_1^\pm}^2
   s_{\beta_{2}}^2\right)\right)+s_{\beta_{1}}^2 \left(c_{\gamma_{2}} s_{\beta_{2}}
   s_{\gamma_{2}}
   \left(m_{H_1^\pm}^2-m_{H_2^\pm}^2\right)+m_{12}^2\right) \right]\, ,
  \\[+2mm]
  \lambda_8=&-\frac{2}{c_{\beta_{1}}
   c_{\beta_{2}} s_{\beta_{2}} v^2} \left[c_{\beta_{1}} c_{\beta_{2}} s_{\beta_{2}}
  \left(c_{\gamma_{2}}^2 m_{H_2^\pm}^2+m_{H_1^\pm}^2
   s_{\gamma_{2}}^2\right)+c_{\beta_{2}} c_{\gamma_{2}} s_{\beta_{1}} s_{\gamma_{2}}
 \left(m_{H_2^\pm}^2-m_{H_1^\pm}^2\right)\right.\nonumber\\
&\left.\hskip 25mm
 +m_{13}^2\right]\, ,
  \\[+2mm]
  \lambda_9=&-\frac{2}{c_{\beta_{2}} s_{\beta_{1}}
    s_{\beta_{2}} v^2}
  \left[c_{\beta_{2}} \left(c_{\beta_{1}} c_{\gamma_{2}}
  s_{\gamma_{2}}
   \left(m_{H_1^\pm}^2-m_{H_2^\pm}^2\right)+c_{\gamma_{2}}^2 m_{H_2^\pm}^2
   s_{\beta_{1}} s_{\beta_{2}}+m_{H_1^\pm}^2 s_{\beta_{1}} s_{\beta_{2}}
   s_{\gamma_{2}}^2\right)\right.\nonumber\\
&\left.\hskip 25mm
 +m_{23}^2\right]\, ,
  \\[+2mm]
  m^2_{12}=&c_{\beta_{1}}^2 c_{\gamma_{1}} s_{\beta_{2}} s_{\gamma_{1}}
   \left(m_{A_{1}}^2-m_{A_{2}}^2\right)+c_{\beta_{1}} s_{\beta_{1}}
   \left(c_{\gamma_{1}}^2 \left(m_{A_{2}}^2
   s_{\beta_{2}}^2-m_{A_{1}}^2\right)+s_{\gamma_{1}}^2 \left(m_{A_{1}}^2
   s_{\beta_{2}}^2-m_{A_{2}}^2\right)\right)\nonumber\\
&\hskip 2mm
+c_{\gamma_{1}} s_{\beta_{1}}^2
   s_{\beta_{2}} s_{\gamma_{1}}
   \left(m_{A_{2}}^2-m_{A_{1}}^2\right)\, ,
  \\[+2mm]
  m^2_{13}=&-c_{\beta_{2}} \left(c_{\beta_{1}} s_{\beta_{2}} \left(c_{\gamma_{1}}^2
  m_{A_{2}}^2+m_{A_{1}}^2
   s_{\gamma_{1}}^2\right)+c_{\gamma_{1}} s_{\beta_{1}} s_{\gamma_{1}}
   \left(m_{A_{2}}^2-m_{A_{1}}^2\right)\right)\, ,
  \\[+2mm]
  m^2_{23}=&-c_{\beta_{2}} \left(c_{\beta_{1}} c_{\gamma_{1}} s_{\gamma_{1}}
   \left(m_{A_{1}}^2-m_{A_{2}}^2\right)+c_{\gamma_{1}}^2 m_{A_{2}}^2
   s_{\beta_{1}} s_{\beta_{2}}+m_{A_{1}}^2 s_{\beta_{1}} s_{\beta_{2}}
   s_{\gamma_{1}}^2\right)\, .
\end{align}

\subsection{The $U(1)\times\Z2$ potential}

As there are no $\lambda''_{11}$ and $\lambda''_{12}$,
we can also solve for two of the soft terms. We choose
$m^2_{13},m^2_{23}$ leaving $m^2_{12}$ as independent. The expressions are
\begin{align}
\label{eq:42}
\lambda_1=&\frac{1}{2 c_{\beta_{1}}^3 c_{\beta_{2}}^3 v^2}
\left[c_{\alpha_{1}}^2 c_{\beta_{1}} c_{\beta_{2}}
  \left(c_{\alpha_{2}}^2
   m_{h}^2+s_{\alpha_{2}}^2 \left(c_{\alpha_{3}}^2
     m_{H_{2}}^2+m_{H_{1}}^2 s_{\alpha_{3}}^2\right)\right)\right.\nonumber\\
&\left.\hskip 20mm
 +2 c_{\alpha_{1}}
   c_{\alpha_{3}} c_{\beta_{1}} c_{\beta_{2}} s_{\alpha_{1}} s_{\alpha_{2}} s_{\alpha_{3}}
   \left(m_{H_{1}}^2-m_{H_{2}}^2\right)+c_{\alpha_{3}}^2 c_{\beta_{1}}
   c_{\beta_{2}} m_{H_{1}}^2 s_{\alpha_{1}}^2\right.\nonumber\\
&\left.\hskip 20mm
   +c_{\beta_{1}} c_{\beta_{2}}
   m_{H_{2}}^2 s_{\alpha_{1}}^2 s_{\alpha_{3}}^2+c_{\beta_{2}} m_{12}^2
   s_{\beta_{1}}+m_{13}^2 s_{\beta_{2}}\right]\, ,
  \\[+2mm]
  \lambda_2=&\frac{1}{2 c_{\beta_{2}}^3
    s_{\beta_{1}}^3 v^2}
  \left[c_{\alpha_{1}}^2 c_{\beta_{2}} s_{\beta_{1}} \left(c_{\alpha_{3}}^2
   m_{H_{1}}^2+m_{H_{2}}^2 s_{\alpha_{3}}^2\right)+2 c_{\alpha_{1}} c_{\alpha_{3}}
   c_{\beta_{2}} s_{\alpha_{1}} s_{\alpha_{2}} s_{\alpha_{3}} s_{\beta_{1}}
   \left(m_{H_{2}}^2-m_{H_{1}}^2\right)\right.\nonumber\\
 &\left.\hskip 20mm
   +c_{\alpha_{2}}^2 c_{\beta_{2}}
   m_{h}^2 s_{\alpha_{1}}^2 s_{\beta_{1}}+c_{\alpha_{3}}^2 c_{\beta_{2}}
   m_{H_{2}}^2 s_{\alpha_{1}}^2 s_{\alpha_{2}}^2 s_{\beta_{1}}+c_{\beta_{1}} c_{\beta_{2}}
   m_{12}^2\right.\nonumber\\
 &\left.\hskip 20mm
   +c_{\beta_{2}} m_{H_{1}}^2 s_{\alpha_{1}}^2 s_{\alpha_{2}}^2
   s_{\alpha_{3}}^2 s_{\beta_{1}}+m_{23}^2 s_{\beta_{2}}\right]\, ,
  \\[+2mm]
  \lambda_3=&\frac{1}{2 s_{\beta_{2}}^3 v^2}
  \left[c_{\alpha_{2}}^2 c_{\alpha_{3}}^2 m_{H_{2}}^2 s_{\beta_{2}}+c_{\alpha_{2}}^2
   m_{H_{1}}^2 s_{\alpha_{3}}^2 s_{\beta_{2}}+c_{\beta_{1}} c_{\beta_{2}}
   m_{13}^2+c_{\beta_{2}} m_{23}^2 s_{\beta_{1}}+m_{h}^2
   s_{\alpha_{2}}^2 s_{\beta_{2}}\right]\, ,
  \\[+2mm]
  \lambda_4=&\frac{1}{c_{\beta_{1}} c_{\beta_{2}}^2 s_{\beta_{1}} v^2}
  \left[c_{\alpha_{1}}^2 c_{\alpha_{3}} s_{\alpha_{2}} s_{\alpha_{3}}
   \left(m_{H_{2}}^2-m_{H_{1}}^2\right)+c_{\alpha_{1}} s_{\alpha_{1}}
   \left(c_{\alpha_{2}}^2 m_{h}^2+c_{\alpha_{3}}^2 \left(m_{H_{2}}^2
       s_{\alpha_{2}}^2-m_{H_{1}}^2\right)\right.\right.\nonumber\\
 &\left.\left.\hskip 20mm
     +s_{\alpha_{3}}^2 \left(m_{H_{1}}^2
   s_{\alpha_{2}}^2-m_{H_{2}}^2\right)\right)+c_{\alpha_{3}} m_{H_{1}}^2
   s_{\alpha_{1}}^2 s_{\alpha_{2}} s_{\alpha_{3}}-c_{\alpha_{3}} m_{H_{2}}^2 s_{\alpha_{1}}^2
   s_{\alpha_{2}} s_{\alpha_{3}}\right.\nonumber\\
 &\left.\hskip 20mm
   -c_{\beta_{1}} c_{\beta_{2}}^2 \lambda_{7} s_{\beta_{1}} v^2-2
   c_{\beta_{1}} c_{\beta_{2}}^2 \lambda''_{10} s_{\beta_{1}}
   v^2-m_{12}^2 \right]\, ,
  \\[+2mm]
  \lambda_5=&-\frac{1}{c_{\beta_{1}} c_{\beta_{2}} s_{\beta_{2}} v^2}
  \left[c_{\alpha_{1}} c_{\alpha_{2}} s_{\alpha_{2}} \left(c_{\alpha_{3}}^2
   m_{H_{2}}^2-m_{h}^2+m_{H_{1}}^2 s_{\alpha_{3}}^2\right)+c_{\alpha_{2}}
 c_{\alpha_{3}} m_{H_{1}}^2 s_{\alpha_{1}} s_{\alpha_{3}}\right.\nonumber\\
&\left.\hskip 25mm
 -c_{\alpha_{2}} c_{\alpha_{3}}
   m_{H_{2}}^2 s_{\alpha_{1}} s_{\alpha_{3}}+c_{\beta_{1}} c_{\beta_{2}} \lambda_{8}
   s_{\beta_{2}} v^2+m_{13}^2 \right]\, ,
  \\[+2mm]
  \lambda_6=&-\frac{1}{c_{\beta_{2}} s_{\beta_{1}} s_{\beta_{2}}
    v^2}
  \left[c_{\alpha_{2}} \left(c_{\alpha_{1}} c_{\alpha_{3}} s_{\alpha_{3}}
   \left(m_{H_{2}}^2-m_{H_{1}}^2\right)+c_{\alpha_{3}}^2 m_{H_{2}}^2
   s_{\alpha_{1}} s_{\alpha_{2}}+m_{h}^2 (-s_{\alpha_{1}})
   s_{\alpha_{2}}\right.\right.\nonumber\\
&\left.\left.\hskip 25mm
   +m_{H_{1}}^2
   s_{\alpha_{1}} s_{\alpha_{2}} s_{\alpha_{3}}^2\right)+c_{\beta_{2}} \lambda_{9}
   s_{\beta_{1}} s_{\beta_{2}} v^2+m_{23}^2 \right]\, ,
  \\[+2mm]
  \lambda_7=&-\frac{2}{
    c_{\beta_{1}} c_{\beta_{2}}^2 s_{\beta_{1}} v^2}
  \left[c_{\beta_{1}}^3 c_{\beta_{2}}^2 \lambda''_{10} s_{\beta_{1}}
   v^2+c_{\beta_{1}}^2 \left(c_{\gamma_{2}} s_{\beta_{2}} s_{\gamma_{2}}
     \left(m_{H_2^\pm}^2-m_{H_1^\pm}^2\right)+m_{12}^2\right)
 \right.\nonumber\\ 
&\left.\hskip 25mm
   +c_{\beta_{1}}
   s_{\beta_{1}} \left(c_{\beta_{2}}^2 \lambda''_{10} s_{\beta_{1}}^2 v^2+c_{\gamma_{2}}^2
   \left(m_{H_1^\pm}^2-m_{H_2^\pm}^2 s_{\beta_{2}}^2\right)-m_{H_1^\pm}^2
   s_{\beta_{2}}^2 s_{\gamma_{2}}^2+m_{H_2^\pm}^2
   s_{\gamma_{2}}^2\right) \right.\nonumber\\ 
&\left.\hskip 25mm
 +s_{\beta_{1}}^2
   \left(c_{\gamma_{2}} s_{\beta_{2}} s_{\gamma_{2}}
   \left(m_{H_1^\pm}^2-m_{H_2^\pm}^2\right)+m_{12}^2\right)\right]\, ,
  \\[+2mm]
  \lambda_8=&-\frac{2}{c_{\beta_{1}}
    c_{\beta_{2}} s_{\beta_{2}} v^2}
  \left[c_{\beta_{1}} c_{\beta_{2}} s_{\beta_{2}} \left(c_{\gamma_{2}}^2
   m_{H_2^\pm}^2+m_{H_1^\pm}^2 s_{\gamma_{2}}^2\right)+c_{\beta_{2}} c_{\gamma_{2}}
   s_{\beta_{1}} s_{\gamma_{2}}
   \left(m_{H_2^\pm}^2-m_{H_1^\pm}^2\right)+m_{13}^2\right]\, ,
  \\[+2mm]
  \lambda_9=&-\frac{2}{c_{\beta_{2}} s_{\beta_{1}}
    s_{\beta_{2}} v^2}
  \left[c_{\beta_{2}} \left(c_{\beta_{1}} c_{\gamma_{2}} s_{\gamma_{2z}}
   \left(m_{H_1^\pm}^2-m_{H_2^\pm}^2\right)+c_{\gamma_{2}}^2 m_{H_2^\pm}^2
   s_{\beta_{1}} s_{\beta_{2}}+m_{H_1^\pm}^2 s_{\beta_{1}} s_{\beta_{2}}
   s_{\gamma_{2}}^2\right)+m_{23}^2\right]\, ,
  \\[+2mm]
  \lambda''_{10}=&-\frac{1}{2
    c_{\beta_{1}} c_{\beta_{2}}^2 s_{\beta_{1}} v^2}
  \left[c_{\beta_{1}}^2 \left(c_{\gamma_{1}} s_{\beta_{2}} s_{\gamma_{1}}
   \left(m_{A_{2}}^2-m_{A_{1}}^2\right)+m_{12}^2\right)+c_{\beta_{1}}
   s_{\beta_{1}} \left(c_{\gamma_{1}}^2 \left(m_{A_{1}}^2-m_{A_{2}}^2
       s_{\beta_{2}}^2\right)\right.\right.\nonumber\\
&\left.\left.\hskip 25mm
     +s_{\gamma_{1}}^2 \left(m_{A_{2}}^2-m_{A_{1}}^2
   s_{\beta_{2}}^2\right)\right)+s_{\beta_{1}}^2 \left(c_{\gamma_{1}} s_{\beta_{2}}
   s_{\gamma_{1}} \left(m_{A_{1}}^2-m_{A_{2}}^2\right)+m_{12}^2\right)\right]\, ,
  \\[+2mm]
  m^2_{13}=&-c_{\beta_{2}} \left[c_{\beta_{1}} s_{\beta_{2}} \left(c_{\gamma_{1}}^2
   m_{A_{2}}^2+m_{A_{1}}^2 s_{\gamma_{1}}^2\right)+c_{\gamma_{1}} s_{\beta_{1}}
 s_{\gamma_{1}} \left(m_{A_{2}}^2-m_{A_{1}}^2\right)\right]\, ,
\\[+2mm]
  m^2_{23}=&-c_{\beta_{2}} \left[c_{\beta_{1}} c_{\gamma_{1}} s_{\gamma_{1}}
   \left(m_{A_{1}}^2-m_{A_{2}}^2\right)+c_{\gamma_{1}}^2 m_{A_{2}}^2
   s_{\beta_{1}} s_{\beta_{2}}+m_{A_{1}}^2 s_{\beta_{1}} s_{\beta_{2}}
   s_{\gamma_{1}}^2\right]\, .
\end{align}

\subsection{The $\Z2\times\Z2$ potential}

We have
\begin{align}
  \lambda_{1}=&
  \frac{1}{2 c_{\beta_{1}}^3
    c_{\beta_{2}}^3 v^2}
  \left[c_{\alpha_{1}}^2 c_{\beta_{1}} c_{\beta_{2}}
    \left(c_{\alpha_{2}}^2 m_{h}^2+s_{\alpha_{2}}^2 
   \left(c_{\alpha_{3}}^2 m_{H_{2}}^2+m_{H_{1}}^2
     s_{\alpha_{3}}^2\right)\right)\right.\nonumber\\
 &\left.\hskip 20mm
 +2 c_{\alpha_{1}} 
   c_{\alpha_{3}} c_{\beta_{1}} c_{\beta_{2}} s_{\alpha_{1}}
   s_{\alpha_{2}} s_{\alpha_{3}} 
   \left(m_{H_{1}}^2-m_{H_{2}}^2\right)
   +c_{\alpha_{3}}^2 c_{\beta_{1}}
   c_{\beta_{2}} m_{H_{1}}^2 
   s_{\alpha_{1}}^2\right.\nonumber\\
 &\left.\hskip 20mm
   +c_{\beta_{1}} c_{\beta_{2}} m_{H_{2}}^2
   s_{\alpha_{1}}^2 s_{\alpha_{3}}^2+c_{\beta_{2}} 
   m^2_{12} s_{\beta_{1}}+m^2_{13} s_{\beta_{2}}\right]\, ,
\\[+2mm]
 \lambda_{2}=&
\frac{1}{2 c_{\beta_{2}}^3 s_{\beta_{1}}^3 
   v^2}\left[c_{\alpha_{1}}^2 c_{\beta_{2}} s_{\beta_{1}}
  \left(c_{\alpha_{3}}^2 m_{H_{1}}^2+m_{H_{2}}^2 
   s_{\alpha_{3}}^2\right)+2 c_{\alpha_{1}} c_{\alpha_{3}}
 c_{\beta_{2}} s_{\alpha_{1}} s_{\alpha_{2}} s_{\alpha_{3}} 
 s_{\beta_{1}} \left(m_{H_{2}}^2-m_{H_{1}}^2\right) \right. \nonumber\\
&\left.\hskip 20mm
 +c_{\alpha_{2}}^2
   c_{\beta_{2}} m_{h}^2 
   s_{\alpha_{1}}^2 s_{\beta_{1}}+c_{\alpha_{3}}^2 c_{\beta_{2}}
   m_{H_{2}}^2 s_{\alpha_{1}}^2 s_{\alpha_{2}}^2 
   s_{\beta_{1}}\right. \nonumber\\
&\left.\hskip 20mm
   +c_{\beta_{1}} c_{\beta_{2}} m^2_{12}+c_{\beta_{2}}
   m_{H_{1}}^2 s_{\alpha_{1}}^2 
   s_{\alpha_{2}}^2 s_{\alpha_{3}}^2 s_{\beta_{1}}+m^2_{23}
   s_{\beta_{2}} \right]\, ,
\\[+2mm]
 \lambda_{3}=&
 \frac{1}{2 s_{\beta_{2}}^3 v^2}
 \left[c_{\alpha_{2}}^2 c_{\alpha_{3}}^2 m_{H_{2}}^2
  s_{\beta_{2}}+c_{\alpha_{2}}^2 m_{H_{1}}^2 s_{\alpha_{3}}^2 
   s_{\beta_{2}}+c_{\beta_{1}} c_{\beta_{2}} m^2_{13}+c_{\beta_{2}} m^2_{23}
   s_{\beta_{1}}+m_{h}^2 s_{\alpha_{2}}^2 s_{\beta_{2}}\right]\, ,
 \\[+2mm]
 \lambda_{4}=&
\frac{1}{c_{\beta_{1}}
  c_{\beta_{2}}^2 s_{\beta_{1}} v^2}
\left[c_{\alpha_{1}}^2 c_{\alpha_{3}} s_{\alpha_{2}} s_{\alpha_{3}}
   \left(m_{H_{2}}^2-m_{H_{1}}^2\right)+c_{\alpha_{1}} s_{\alpha_{1}}
   \left(c_{\alpha_{2}}^2 
   m_{h}^2+c_{\alpha_{3}}^2 \left(m_{H_{2}}^2
     s_{\alpha_{2}}^2-m_{H_{1}}^2\right)\right.\right.\nonumber\\
&\left.\left.\hskip 20mm
   +s_{\alpha_{3}}^2 
   \left(m_{H_{1}}^2
     s_{\alpha_{2}}^2-m_{H_{2}}^2\right)\right)+c_{\alpha_{3}}
 m_{H_{1}}^2 
   s_{\alpha_{1}}^2 s_{\alpha_{2}} s_{\alpha_{3}}-c_{\alpha_{3}}
   m_{H_{2}}^2 s_{\alpha_{1}}^2 s_{\alpha_{2}} 
   s_{\alpha_{3}}\right.\nonumber\\
 &\left.\hskip 20mm
   -c_{\beta_{1}} c_{\beta_{2}}^2 \lambda_{7}
   s_{\beta_{1}} v^2
   -2 c_{\beta_{1}} c_{\beta_{2}}^2 
   \lambda''_{10} s_{\beta_{1}} v^2-m^2_{12}\right]\, ,
 \\[+2mm]
 \lambda_{5}=&
-\frac{1}{c_{\beta_{1}}
  c_{\beta_{2}} s_{\beta_{2}} v^2}
\left[c_{\alpha_{1}} c_{\alpha_{2}} s_{\alpha_{2}} \left(c_{\alpha_{3}}^2
   m_{H_{2}}^2-m_{h}^2+m_{H_{1}}^2
   s_{\alpha_{3}}^2\right)+c_{\alpha_{2}} c_{\alpha_{3}} 
 m_{H_{1}}^2 s_{\alpha_{1}} s_{\alpha_{3}}\right.\nonumber\\
&\left.\hskip 25mm
 -c_{\alpha_{2}}
   c_{\alpha_{3}} m_{H_{2}}^2 s_{\alpha_{1}} 
   s_{\alpha_{3}}+c_{\beta_{1}} c_{\beta_{2}} \lambda_{8}
   s_{\beta_{2}} v^2+2 c_{\beta_{1}} c_{\beta_{2}} 
   \lambda''_{11} s_{\beta_{2}} v^2+m^2_{13}\right]\, ,
 \\[+2mm]
 \lambda_{6}=&
 -\frac{1}{c_{\beta_{2}} s_{\beta_{1}} s_{\beta_{2}} v^2}
 \left[c_{\alpha_{2}} \left(c_{\alpha_{1}} c_{\alpha_{3}} s_{\alpha_{3}}
   \left(m_{H_{2}}^2-m_{H_{1}}^2\right)+c_{\alpha_{3}}^2 m_{H_{2}}^2
   s_{\alpha_{1}} 
   s_{\alpha_{2}}+m_{h}^2 (-s_{\alpha_{1}}) s_{\alpha_{2}}\right.\right.
 \nonumber\\
&\left.\left. \hskip 25mm
   +m_{H_{1}}^2
   s_{\alpha_{1}} s_{\alpha_{2}} 
   s_{\alpha_{3}}^2\right)+c_{\beta_{2}} s_{\beta_{1}} s_{\beta_{2}}
 v^2 (\lambda_{9}+2 
   \lambda''_{12})+m^2_{23}\right]\, ,
 \\[+2mm]
 \lambda_{7}=&
-\frac{2}{c_{\beta_{1}} 
  c_{\beta_{2}}^2 s_{\beta_{1}} v^2}
\left[c_{\beta_{1}}^3 c_{\beta_{2}}^2 \lambda''_{10}
    s_{\beta_{1}} v^2+c_{\beta_{1}}^2 
   \left(c_{\gamma_{2}} s_{\beta_{2}} s_{\gamma_{2}}
     \left(m_{H_2^\pm}^2-m_{H_1^\pm}^2\right)+m^2_{12}\right)\right.
 \nonumber\\
 &\left.\hskip 25mm
   +c_{\beta_{1}}
 s_{\beta_{1}} 
   \left(c_{\beta_{2}}^2 \lambda''_{10} s_{\beta_{1}}^2 v^2+c_{\gamma_{2}}^2
   \left(m_{H_1^\pm}^2-m_{H_2^\pm}^2 s_{\beta_{2}}^2\right)-m_{H_1^\pm}^2
   s_{\beta_{2}}^2 
   s_{\gamma_{2}}^2+m_{H_2^\pm}^2
   s_{\gamma_{2}}^2\right)\right.\nonumber\\
&\left.\hskip 25mm
 +s_{\beta_{1}}^2 \left(c_{\gamma_{2}}
   s_{\beta_{2}} 
   s_{\gamma_{2}}
   \left(m_{H_1^\pm}^2-m_{H_2^\pm}^2\right)+m^2_{12}\right)\right]\, ,
 \\[+2mm]
 \lambda_{8}=&
-\frac{2}{c_{\beta_{1}}
  c_{\beta_{2}}    s_{\beta_{2}} v^2}
\left[c_{\beta_{1}} c_{\beta_{2}} s_{\beta_{2}}
    \left(c_{\gamma_{2}}^2 m_{H_2^\pm}^2+\lambda''_{11} 
   v^2+m_{H_1^\pm}^2 s_{\gamma_{2}}^2\right)+c_{\beta_{2}}
 c_{\gamma_{2}} s_{\beta_{1}} s_{\gamma_{2}} 
   \left(m_{H_2^\pm}^2-m_{H_1^\pm}^2\right)\right.\nonumber\\
 &\left.\hskip 25mm
   +m^2_{13}\right]\, ,
 \\[+2mm]
 \lambda_{9}=&
-\frac{2}{c_{\beta_{2}}
  s_{\beta_{1}} s_{\beta_{2}} v^2}
\left[c_{\beta_{2}} \left(c_{\beta_{1}} c_{\gamma_{2}} s_{\gamma_{2}}
   \left(m_{H_1^\pm}^2-m_{H_2^\pm}^2\right)+c_{\gamma_{2}}^2 m_{H_2^\pm}^2
   s_{\beta_{1}} 
   s_{\beta_{2}}\right.\right.\nonumber\\
 &\left.\left.\hskip 25mm
   +s_{\beta_{1}} s_{\beta_{2}} \left(\lambda''_{12} v^2+m_{H_1^\pm}^2
   s_{\gamma_{2}}^2\right)\right)+m^2_{23}\right]\, ,
 \\[+2mm]
 \lambda''_{10}=&
-\frac{1}{2  c_{\beta_{1}} c_{\beta_{2}}^2 s_{\beta_{1}} v^2}
\left[c_{\beta_{1}}^2 \left(c_{\gamma_{1}} s_{\beta_{2}} s_{\gamma_{1}}
   \left(m_{A_{2}}^2-m_{A_{1}}^2\right)+m^2_{12}\right)+c_{\beta_{1}}
 s_{\beta_{1}} 
   \left(c_{\gamma_{1}}^2 \left(m_{A_{1}}^2-m_{A_{2}}^2
       s_{\beta_{2}}^2\right)\right.\right.\nonumber\\
&\left.\left.\hskip 25mm
     +s_{\gamma_{1}}^2 
   \left(m_{A_{2}}^2-m_{A_{1}}^2
     s_{\beta_{2}}^2\right)\right)+s_{\beta_{1}}^2
 \left(c_{\gamma_{1}} 
   s_{\beta_{2}} s_{\gamma_{1}}
   \left(m_{A_{1}}^2-m_{A_{2}}^2\right)+m^2_{12}\right)\right] \, ,
 \\[+2mm]
 \lambda''_{11}=&
-\frac{c_{\beta_{1}} c_{\beta_{2}} s_{\beta_{2}}
  \left(c_{\gamma_{1}}^2 m_{A_{2}}^2+m_{A_{1}}^2 
   s_{\gamma_{1}}^2\right)+c_{\beta_{2}} c_{\gamma_{1}} s_{\beta_{1}}
 s_{\gamma_{1}} 
   \left(m_{A_{2}}^2-m_{A_{1}}^2\right)+m^2_{13}}{2 c_{\beta_{1}}
   c_{\beta_{2}} s_{\beta_{2}}
   v^2}\, ,
 \\[+2mm]
 \lambda''_{12}=&
 -\frac{c_{\beta_{2}} \left(c_{\beta_{1}} c_{\gamma_{1}} s_{\gamma_{1}}
   \left(m_{A_{1}}^2-m_{A_{2}}^2\right)+c_{\gamma_{1}}^2 m_{A_{2}}^2
   s_{\beta_{1}} 
   s_{\beta_{2}}+m_{A_{1}}^2 s_{\beta_{1}} s_{\beta_{2}}
   s_{\gamma_{1}}^2\right)+m^2_{23}}{2 
   c_{\beta_{2}} s_{\beta_{1}} s_{\beta_{2}} v^2}\, .
\end{align}

\section{\label{app:STU}Oblique parameters $STU$}

In order to discuss the effect of the $S,T,U$ parameters,
we use the results in \cite{Grimus:2007if}.
To apply the relevant expressions,
we write the matrices $U$ and $V$ used in \cite{Grimus:2007if}
with the notation choices that we made when obtaining the mass
eigenstates in Section~\ref{sec:scalar}.
We start with the $3\times 6$ matrix $V$ defined as
  \begin{equation}
      \begin{pmatrix}x_1+\,i\,z_1 \\ x_2+\,i\,z_2 \\ x_3+\,i\,z_3  \end{pmatrix} = V \begin{pmatrix}G^0 \\ h_1 \\ h_2 \\ h_3 \\ A_1 \\ A_2 \end{pmatrix} ,
  \end{equation}
and find, by comparison with \eqs{CPevenDiag}{CPoddDiag}, that $V$ is 
 \begin{equation}
     V= \begin{pmatrix}[1.5] i \textbf{P}^T_{11} & \textbf{R}^T_{11}& \textbf{R}^T_{12}&\textbf{R}^T_{13} &  i \textbf{P}^T_{12} &  i \textbf{P}^T_{13} \\ i \textbf{P}^T_{21} & \textbf{R}^T_{21}& \textbf{R}^T_{22}&\textbf{R}^T_{23} &  i \textbf{P}^T_{22} &  i \textbf{P}^T_{23}\\i \textbf{P}^T_{31} & \textbf{R}^T_{31}& \textbf{R}^T_{32}&\textbf{R}^T_{33} &  i \textbf{P}^T_{32} &  i \textbf{P}^T_{33}  \end{pmatrix} .
 \end{equation}
  
The $3\times 3$ matrix U defined as
\begin{equation}
\begin{pmatrix} w_1^\dagger \\ w_2^\dagger \\ w_3^\dagger \end{pmatrix}
=U \begin{pmatrix} G^\dagger \\ H_1^+ \\ H_2^+  \end{pmatrix},
\label{chargedtransfU}
\end{equation}
gives us the correspondence $U=\textbf{Q}^T $ from \eq{ChargedDiag}.
  
Having applied the expressions for $S, T, U $,
the constraints implemented on $S$ and $T$ follow
Ref. \cite{Baak:2014ora}, at $95\%$ confidence level.
For $U$, we fix the allowed interval to be
\begin{equation}
    U=0.03\pm0.10 .
\end{equation}

\section{\label{app:Unitarity}Perturbative Unitarity  Constraints}

In order to determine the tree-level unitarity constraints, we use
the algorithm presented in \cite{Bento:2017eti}. As described there,
we have to impose that the eigenvalues of the scattering S-matrix of
two scalars into two scalars have an upper bound (the unitarity
limit).
We separate the scattering matrices according to charge and
hypercharge \cite{Bento:2017eti}, $M_{2Y}^Q$. For our models this has
been done in Ref.~\cite{Bento:2022vsb} and we copy here their results.
We consider the case of $\Z2\times \Z2$, as the others can be obtained
from this by setting some of the $\lambda'$s to zero.

\subsubsection{$M_2^{++}$}
We get,
\begin{equation}
  \label{eq:8b}
  M_2^{++}=\text{diag}\left\{
    \begin{bmatrix}
      2\lambda_1&2\lambda'_{10}&2\lambda'_{11}\\
      2\lambda'_{10}&2\lambda_{2}&2\lambda'_{12}\\
     2\lambda'_{11} &2\lambda'_{12}&2\lambda_{3}
   \end{bmatrix},(\lambda_4+\lambda_7),
   (\lambda_5+\lambda_8),
   (\lambda_6+\lambda_9)\right\}.
\end{equation}

\subsubsection{$M_2^{+}$}
We get
\begin{equation}
  \label{eq:9b}
  M_2^{+}= \text{diag}\left\{
    \begin{bmatrix}
      2\lambda_1&2\lambda'_{10}&2\lambda'_{11}\\
      2\lambda'_{10}&2\lambda_{2}&2\lambda'_{12}\\
     2\lambda'_{11} &2\lambda'_{12}&2\lambda_{3}
   \end{bmatrix},
   \begin{bmatrix}
     \lambda_4&\lambda_7\\
     \lambda_7&\lambda_4
   \end{bmatrix},
      \begin{bmatrix}
     \lambda_5&\lambda_8\\
     \lambda_8&\lambda_5
   \end{bmatrix},
      \begin{bmatrix}
     \lambda_6&\lambda_9\\
     \lambda_9&\lambda_6
   \end{bmatrix}
    \right\}.
\end{equation}

\subsubsection{$M_0^{+}$}
We get,
\begin{equation}
  \label{eq:14b}
  M_0^{+}= \text{diag}\left\{
    \begin{bmatrix}
      2\lambda_1&\lambda_{7}&\lambda_{8}\\
      \lambda_{7}&2\lambda_{2}&\lambda_{9}\\
     \lambda_{8} &\lambda_{9}&2\lambda_{3}
   \end{bmatrix},
   \begin{bmatrix}
     \lambda_4&2\lambda'_{10}\\
     2\lambda'_{10}&\lambda_4
   \end{bmatrix},
      \begin{bmatrix}
     \lambda_5&2\lambda'_{11}\\
     2\lambda'_{11}&\lambda_5
   \end{bmatrix},
      \begin{bmatrix}
     \lambda_6&2\lambda'_{12}\\
     2\lambda'_{12}&\lambda_6
   \end{bmatrix}
  \right\}  
\end{equation}

\subsubsection{$M_0^{0}$}
We get,
\begin{align}
  \label{eq:15b}
  M_0^{0}=&\text{diag}\left\{
    M_0^+,
    \begin{bmatrix}
      6\lambda_1&2\lambda_4+\lambda_7&2\lambda_5+\lambda_8\\
      2\lambda_4+\lambda_7&6\lambda_2&2\lambda_6+\lambda_9\\
      2\lambda_5+\lambda_8&2\lambda_6+\lambda_9&6\lambda_3\\
    \end{bmatrix},
    \begin{bmatrix}
      \lambda_4+2\lambda_7&6\lambda'_{10}\\
      6\lambda'_{10}&\lambda_4+2\lambda_7
    \end{bmatrix},
  \right.\nonumber\\[+2mm]
  &\hskip 10mm \left.
    \begin{bmatrix}
      \lambda_5+2\lambda_8&6\lambda'_{11}\\
      6\lambda'_{11}&\lambda_5+2\lambda_8
    \end{bmatrix},\begin{bmatrix}
      \lambda_6+2\lambda_9&6\lambda'_{12}\\
      6\lambda'_{12}&\lambda_6+2\lambda_9
    \end{bmatrix}
    \right\}.
\end{align}
Denoting by $\Lambda_i$ the eigenvalues of the relevant scattering matrices,
we have 27 $\Lambda$'s to calculate for each set of physical
parameters randomly generated, and the condition to impose is that 
\begin{equation}
     |\Lambda_i| \leq 8\pi\,\,,\quad i=1,..,21\, .
\end{equation}
The explicit expressions for the different eigenvalues are,
\begin{align}
  \label{eq:16_2}
  &\Lambda_{1-3}^{++}=\text{Roots of}:\nonumber\\
  &\hskip 10mm x^3 + 2 (-\lambda_1-\lambda_2-\lambda_3)x^2
  +4\left(-\lambda'{}_{10}^2-\lambda'{}_{11}^2-\lambda'{}_{12}^2
    +\lambda_1\lambda_2+\lambda_1\lambda_3+\lambda_2\lambda_3
  \right) x\nonumber\\
  &\hskip 10 mm +8\left(\lambda_3 \lambda'{}_{10}^2+
    \lambda_2 \lambda'{}_{11}^2 +\lambda_1 \lambda'{}_{12}^2
    -2 \lambda'{}_{10}\lambda'{}_{11}\lambda'{}_{12}
    -\lambda_1\lambda_2\lambda_3\right)=0 ,\\
  &\Lambda_{4}^{++}= \lambda_4+\lambda_7 ,\\
  &\Lambda_{5}^{++}= \lambda_5+\lambda_8 ,\\
  &\Lambda_{6}^{++}= \lambda_6+\lambda_9 ,\\
  &\Lambda_{1-3}^{+,2}=\Lambda_{1-3}^{++} , \\
  &\Lambda_{4,5}^{+,2}= \lambda_4\pm\lambda_7 ,\\
  &\Lambda_{6,7}^{+,2}= \lambda_5\pm\lambda_8 ,\\
  &\Lambda_{8,9}^{+,2}= \lambda_6\pm\lambda_9 ,\\
  &\Lambda_{1-3}^{+,0}=\text{Roots of}:\nonumber\\
  &\hskip 10mm x^3 +2 (-\lambda_1-\lambda_2-\lambda_3)x^2
  +\left(-\lambda_7^2-\lambda_8^2-\lambda_9^2
    +4 \lambda_1\lambda_2+4 \lambda_1\lambda_3+4 \lambda_2\lambda_3
  \right) x\nonumber\\
  &\hskip 10mm +2 \left(\lambda_3 \lambda_7^2+\lambda_2
    \lambda_8^2+\lambda_1 \lambda_9^2 -\lambda_7\lambda_8\lambda_9
    -4\lambda_1\lambda_2\lambda_3 \right)=0 ,\\
  &\Lambda_{4,5}^{+,0}= \lambda_4\pm 2\lambda'{}_{10} ,\\
  &\Lambda_{6,7}^{+,0}= \lambda_5\pm 2\lambda'{}_{11} ,\\
  &\Lambda_{8,9}^{+,0}= \lambda_6\pm 2\lambda'{}_{12} ,\\
  &\Lambda_{1-9}^{0,0}=\Lambda_{1-9}^{+,0} , \\
  &\Lambda_{10-12}^{0,0}=\text{Roots of}:\nonumber\\
  &\hskip 10mm x^3 + 6 \left(-\lambda_1-\lambda_2-\lambda_3\right) x^2
  +\left(-4\lambda_4^2-4\lambda_5^2-4\lambda_6^2
    -\lambda_7^2-\lambda_8^2-\lambda_9^2
    -4\lambda_4\lambda_7-4\lambda_5\lambda_8\right.\nonumber\\
  & \hskip 10mm\left.
  -4\lambda_6\lambda_9
    +36\lambda_1\lambda_2+36\lambda_1\lambda_3+36\lambda_2\lambda_3
  \right) x + 2\left( 12 \lambda_3 \lambda_4^2+12 \lambda_2
    \lambda_5^2+12 \lambda_1 \lambda_6^2
    +3\lambda_3\lambda_7^2 \right.\nonumber\\
   & \hskip 10mm\left.
    +3\lambda_2\lambda_8^2+3\lambda_1\lambda_9^2
    -108 \lambda_1\lambda_2\lambda_3
    -8 \lambda_4\lambda_5\lambda_6
    +12\lambda_3\lambda_4\lambda_7
    +12\lambda_2\lambda_5\lambda_8
    +12\lambda_1\lambda_6\lambda_9\right.\\
   & \hskip 10mm\left.
    -4\lambda_5\lambda_6\lambda_7
    -4\lambda_4\lambda_6\lambda_8
    -4\lambda_4\lambda_5\lambda_9
    -2\lambda_6\lambda_7\lambda_8
    -2\lambda_5\lambda_7\lambda_9
    -2\lambda_4\lambda_8\lambda_9
    -\lambda_7\lambda_8\lambda_9\right) =0,\nonumber\\
  &\Lambda_{13-14}^{0,0}=\lambda_4+2\lambda_7\pm 6\lambda'_{10},\\
  &\Lambda_{15-16}^{0,0}=\lambda_5+2\lambda_8\pm 6\lambda'_{11},\\
  &\Lambda_{17-18}^{0,0}=\lambda_6+2\lambda_9\pm 6\lambda'_{12}.
\end{align}
We can take as independent the set
\begin{equation}
  \label{eq:17_2}
  \Lambda_{1-3}^{++}, \Lambda_{4,5}^{+,2}, \Lambda_{6,7}^{+,2},
  \Lambda_{8,9}^{+,2}, \Lambda_{1-9}^{+,0},\Lambda_{10-18}^{0,0} .  
\end{equation}

Now for the case of $U(1)\times\Z2$ the results are obtained from
those above setting $\lambda'_{11}=\lambda'_{12}=0$,
and for the 
$U(1)\times U(1)$ case we should put
$\lambda'_{10}=\lambda'_{11}=\lambda'_{12} =0$. One can check with
Ref.~\cite{Bento:2022vsb} that this is leads to the corerct results.

\section{\label{app:Types}Yukawa interactions in the mass basis}

\subsection{Type-II}

For this case we assume that under the group,
\begin{equation}
\label{eq:Type-IIa}
n_R\to (+,e^{-i \theta'})\,n_R,\qquad \ell_R\to (+,e^{-i \theta'})\,\ell_R ,
\end{equation}
the other fermion fields remaining unaffected,
where we have used the notation of
Table~\ref{tab:Types}, only altered by using ``+'' for
invariance.\footnote{We used a space instead of ``+''
for an invariance in Table~\ref{tab:Types},
in order not to clutter the notation.}    
Therefore, up quarks couple to $\phi_3$ and down quarks and leptons
couple only to $\phi_2$. With the conventions of
Eq.~(\ref{eq:couplingNeutralFerm}) and
Eq.~(\ref{eq:couplingChargedFerm}) we have,
\begin{align}
a_j^f \to&
\frac{\textbf{R}_{j,2}}{\hat{v_2}},
\qquad\qquad j=1,2,3\qquad \text{for all leptons} ,\nonumber\\[2pt]
b_j^f \to&
\frac{\textbf{P}_{j-2,2}}{\hat{v_2}},
\qquad\quad j=4,5\quad\qquad \text{for all leptons} ,\nonumber\\[2pt]
a_j^f \to&
\frac{\textbf{R}_{j,3}}{\hat{v_3}},
\qquad\qquad j=1,2,3\qquad \text{for all up quarks} ,\nonumber\\[2pt]
b_j^f \to&
-\frac{\textbf{P}_{j-2,3}}{\hat{v_3}},
\quad\quad j=4,5\quad\qquad \text{for all up quarks} ,\nonumber\\[2pt]
a_j^f \to&
\frac{\textbf{R}_{j,2}}{\hat{v_2}},
\qquad\qquad j=1,2,3\qquad \text{for all down quarks} ,\nonumber\\[2pt]
b_j^f \to&
\frac{\textbf{P}_{j-2,2}}{\hat{v_2}},
\qquad\quad j=4,5\quad\qquad \text{for all down quarks} ,
\label{eq:coeffNeutralFerm-Type-II}
\end{align}
and
\begin{equation}
\label{eq:coeffChargedFerm-Type-II}  
\eta_k^{\ell\,L}=-\frac{\textbf{Q}_{k+1,2}}{\hat{v_2}}\,,
\quad\eta_k^{\ell\,R}=0\,,\quad\eta_k^{q\,L}
=-\frac{\textbf{Q}_{k+1,2}}{\hat{v_2}}\,,\quad\eta_k^{q\,R}
=\frac{\textbf{Q}_{k+1,3}}{\hat{v_3}}\,,\quad \text{k=1,2} ,
\end{equation}

\subsection{Type-X}

For this case we assume that under the group,
\begin{equation}
  \label{eq:Type-IIb}
    \ell_R\to (+,e^{-i \theta'})\,\ell_R ,
  \end{equation}
the other fermion fields remaining unaffected, where we have used the notation of
Table~\ref{tab:Types}.   
Therefore up and down quarks couple to $\phi_3$ and leptons
couple only to $\phi_2$. With the conventions of
Eq.~(\ref{eq:couplingNeutralFerm}) and
Eq.~(\ref{eq:couplingChargedFerm}) we have,
\begin{align}
a_j^f \to&
\frac{\textbf{R}_{j,2}}{\hat{v_2}},
\qquad\qquad j=1,2,3\qquad \text{for all leptons} ,\nonumber\\[2pt]
b_j^f \to&
\frac{\textbf{P}_{j-2,2}}{\hat{v_2}},
\qquad\quad j=4,5\quad\qquad \text{for all leptons} ,\nonumber\\[2pt]
a_j^f \to&
\frac{\textbf{R}_{j,3}}{\hat{v_3}},
\qquad\qquad j=1,2,3\qquad \text{for all up quarks} ,\nonumber\\[2pt]
b_j^f \to&
-\frac{\textbf{P}_{j-2,3}}{\hat{v_3}},
\quad\quad j=4,5\quad\qquad \text{for all up quarks} ,\nonumber\\[2pt]
a_j^f \to&
\frac{\textbf{R}_{j,3}}{\hat{v_3}},
\qquad\qquad j=1,2,3\qquad \text{for all down quarks} ,\nonumber\\[2pt]
b_j^f \to&
\frac{\textbf{P}_{j-2,3}}{\hat{v_3}},
\qquad\quad j=4,5\quad\qquad \text{for all down quarks} ,
\label{eq:coeffNeutralFerm-Type-X}
\end{align}
and
\begin{equation}
\label{eq:coeffChargedFerm-Type-X}  
\eta_k^{\ell\,L}=-\frac{\textbf{Q}_{k+1,2}}{\hat{v_2}}\,,
\quad\eta_k^{\ell\,R}=0\,,\quad\eta_k^{q\,L}
=-\frac{\textbf{Q}_{k+1,3}}{\hat{v_3}}\,,\quad\eta_k^{q\,R}
=\frac{\textbf{Q}_{k+1,3}}{\hat{v_3}}\,,\quad \text{k=1,2} ,
\end{equation}

\subsection{Type-Y}

For this case we assume that under the group,
\begin{equation}
  \label{eq:Type-IIc}
    n_R\to (+,e^{-i \theta'})\,n_R,
  \end{equation}
the other fermion fields remaining unaffected, where we have used the notation of
Table~\ref{tab:Types}.
Therefore up quarks and leptons couple to $\phi_3$ and down quarks
couple only to $\phi_2$. With the conventions of
Eq.~(\ref{eq:couplingNeutralFerm} and
Eq.~(\ref{eq:couplingChargedFerm}) we have,
\begin{align}
a_j^f \to&
\frac{\textbf{R}_{j,3}}{\hat{v_3}},
\qquad\qquad j=1,2,3\qquad \text{for all leptons} ,\nonumber\\[2pt]
b_j^f \to&
\frac{\textbf{P}_{j-2,3}}{\hat{v_3}},
\qquad\quad j=4,5\quad\qquad \text{for all leptons} ,\nonumber\\[2pt]
a_j^f \to&
\frac{\textbf{R}_{j,3}}{\hat{v_3}},
\qquad\qquad j=1,2,3\qquad \text{for all up quarks} ,\nonumber\\[2pt]
b_j^f \to&
-\frac{\textbf{P}_{j-2,3}}{\hat{v_3}},
\quad\quad j=4,5\quad\qquad \text{for all up quarks} ,\nonumber\\[2pt]
a_j^f \to&
\frac{\textbf{R}_{j,2}}{\hat{v_2}},
\qquad\qquad j=1,2,3\qquad \text{for all down quarks} ,\nonumber\\[2pt]
b_j^f \to&
\frac{\textbf{P}_{j-2,2}}{\hat{v_2}},
\qquad\quad j=4,5\quad\qquad \text{for all down quarks} ,
\label{eq:coeffNeutralFerm-Type-Y}
\end{align}
and
\begin{equation}
\label{eq:coeffChargedFerm-Type-Y}  
\eta_k^{\ell\,L}=-\frac{\textbf{Q}_{k+1,3}}{\hat{v_3}}\,,
\quad\eta_k^{\ell\,R}=0\,,\quad\eta_k^{q\,L}
=-\frac{\textbf{Q}_{k+1,2}}{\hat{v_2}}\,,
\quad\eta_k^{q\,R}=\frac{\textbf{Q}_{k+1,3}}{\hat{v_3}}\,,\quad \text{k=1,2} ,
\end{equation}

\subsection{Type-Z}

For this case we assume that under the group,
\begin{equation}
  \label{eq:Type-IId}
    n_R\to (+,e^{-i \theta'})\,n_R,\qquad \ell_R\to (e^{-i \theta},+)\,\ell_R ,
  \end{equation}
the other fermion fields remaining unaffected, where we have used the notation of
Table~\ref{tab:Types}.
It follows that the
Yukawa coupling matrices are now restricted: $\phi_1$ only has
interaction terms with the charged leptons, giving them mass; $\phi_3$
and $\phi_2$ are responsible for masses of the up and down type
quarks, respectively.

With the conventions of
Eq.~(\ref{eq:couplingNeutralFerm}) and
Eq.~(\ref{eq:couplingChargedFerm}) we have,
\begin{align}
a_j^f &\to&
\frac{\textbf{R}_{j,1}}{\hat{v_1}},
\qquad\qquad j=1,2,3\qquad \text{for all leptons} ,\nonumber\\[2pt]
b_j^f &\to&
\frac{\textbf{P}_{j-2,1}}{\hat{v_1}},
\qquad\quad j=4,5\quad\qquad \text{for all leptons} ,\nonumber\\[2pt]
a_j^f &\to&
\frac{\textbf{R}_{j,3}}{\hat{v_3}},
\qquad\qquad j=1,2,3\qquad \text{for all up quarks} ,\nonumber\\[2pt]
b_j^f &\to&
-\frac{\textbf{P}_{j-2,3}}{\hat{v_3}},
\quad\quad j=4,5\quad\qquad \text{for all up quarks} ,\nonumber\\[2pt]
a_j^f &\to&
\frac{\textbf{R}_{j,2}}{\hat{v_2}},
\qquad\qquad j=1,2,3\qquad \text{for all down quarks} ,\nonumber\\[2pt]
b_j^f &\to&
\frac{\textbf{P}_{j-2,2}}{\hat{v_2}},
\qquad\quad j=4,5\quad\qquad \text{for all down quarks} ,
\label{coeffNeutralFerm}
\end{align}
where we introduce $\hat{v_i}=v_i/v$, with the vevs in \eq{3hdmvevs}.
Note how the coupling of each type of fermion depends on entries of the
diagonalization matrices in \eqs{matrixR}{matrixP}. 

The charged couplings are
\begin{equation}
\eta_k^{\ell\,L}=-\frac{\textbf{Q}_{k+1,1}}{\hat{v_1}}\,,\quad\eta_k^{\ell\,R}=0\,,
\quad\eta_k^{q\,L}=-\frac{\textbf{Q}_{k+1,2}}{\hat{v_2}}\,,
\quad\eta_k^{q\,R}=\frac{\textbf{Q}_{k+1,3}}{\hat{v_3}}\,,\quad \text{k=1,2} ,
\end{equation}
for leptons and quarks, respectively.

\providecommand{\href}[2]{#2}\begingroup\raggedright\endgroup


\begin{thebibliography}{10}

\bibitem{Faro:2019vcd}
F.~S. Faro and I.~P. Ivanov, {\it {Boundedness from below in the $U(1)\times
  U(1)$ three-Higgs-doublet model}},  {\em Phys. Rev. D} {\bf 100} (2019),
  no.~3 035038, [\href{https://arxiv.org/pdf/1907.01963.pdf}{{\tt 1907.01963}}].

\bibitem{Faro:2019}
F.~Faro, {\it {Some Theoretical Aspects of Multi-Higgs-Doublet Models}},
  https://fenix.tecnico.ulisboa.pt/cursos/meft/dissertacao/283828618790340,
  Master's thesis, IST, Univ. Lisbon, 11 November 2019.

\bibitem{Das:2021oik}
D.~Das, P.~M.~Ferreira, A.~P.~Morais, I.~Padilla-Gay, R.~Pasechnik and J.~P.~Rodrigues,
\textit{A three Higgs doublet model with symmetry-suppressed
flavour changing neutral currents},
{\em JHEP}  \textbf{11} (2021), 079
doi:10.1007/JHEP11(2021)079
[arXiv:2106.06425 [hep-ph]].

\bibitem{Weinberg:1976hu}
S.~Weinberg,
\textit{Gauge Theory of CP Violation},
{\em Phys. Rev. Lett.} \textbf{37} (1976), 657.

\bibitem{Grzadkowski:2009bt}
B.~Grzadkowski, O.~M. Ogreid, and P.~Osland, {\it {Natural Multi-Higgs Model
  with Dark Matter and CP Violation}},  {\em Phys. Rev. D} {\bf 80} (2009)
  055013, [\href{https://arxiv.org/pdf/0904.2173.pdf}{{\tt 0904.2173}}].

\bibitem{Klimenko:1984qx}
K.~G. Klimenko, {\it {On Necessary and Sufficient Conditions for Some Higgs
  Potentials to Be Bounded From Below}},  {\em Theor. Math. Phys.} {\bf 62}
  (1985) 58--65. [Teor. Mat. Fiz.62,87(1985)].

\bibitem{Kannike:2012pe}
K.~Kannike, {\it {Vacuum Stability Conditions From Copositivity Criteria}},
  {\em Eur. Phys. J. C} {\bf 72} (2012) 2093,
  [\href{https://arxiv.org/pdf/1205.3781.pdf}{{\tt 1205.3781}}].

\bibitem{Carrolo:2022oyg}
S.~Carrolo, J.~C.~Rom\~{a}o, and J.~P.~Silva,
{\it {Conditions for global minimum in the $A_4$ symmetric 3HDM}}, 
[arXiv:2207.02928 [hep-ph]].


\bibitem{Hernandez-Sanchez:2020aop}
J.~Hernandez-Sanchez, V.~Keus, S.~Moretti, D.~Rojas-Ciofalo, and D.~Sokolowska,
  {\it {Complementary Probes of Two-component Dark Matter}},
  \href{https://arxiv.org/pdf/2012.11621.pdf}{{\tt 2012.11621}}.

\bibitem{Fontes:2014xva}
D.~Fontes, J.~C. Rom\~{a}o, and J.~P. Silva, {\it {$h \rightarrow Z \gamma$ in the
  complex two Higgs doublet model}},  {\em JHEP} {\bf 12} (2014) 043,
  [\href{https://arxiv.org/pdf/1408.2534.pdf}{{\tt 1408.2534}}].

\bibitem{Fontes:2017zfn}
D.~Fontes, M.~M\"{u}hlleitner, J.~C. Rom\~{a}o, R.~Santos, J.~P. Silva, and
  J.~Wittbrodt, {\it {The C2HDM revisited}},  {\em JHEP} {\bf 02} (2018) 073,
  [\href{https://arxiv.org/pdf/1711.09419.pdf}{{\tt 1711.09419}}].

\bibitem{Florentino:2021ybj}
R.~R. Florentino, J.~C. Rom\~{a}o, and J.~P. Silva, {\it {Off diagonal charged
  scalar couplings with the Z boson: Zee-type models as an example}},  {\em
  Eur. Phys. J. C} {\bf 81} (2021), no.~12 1148,
  [\href{https://arxiv.org/pdf/2106.08332.pdf}{{\tt 2106.08332}}].

\bibitem{Boto:2021qgu}
R.~Boto, J.~C. Rom\~{a}o, and J.~P. Silva, {\it {Current bounds on the type-Z
  Z3 three-Higgs-doublet model}},  {\em Phys. Rev. D} {\bf 104} (2021), no.~9
  095006, [\href{https://arxiv.org/pdf/2106.11977.pdf}{{\tt 2106.11977}}].

\bibitem{Fontes:2019wqh}
D.~Fontes and J.~C. Romao, {\it {FeynMaster: a plethora of Feynman tools}},
  {\em Comput. Phys. Commun.} {\bf 256} (2020) 107311,
  [\href{https://arxiv.org/pdf/1909.05876.pdf}{{\tt 1909.05876}}].

\bibitem{Fontes:2021iue}
D.~Fontes and J.~C. Rom\~{a}o, {\it {Renormalization of the C2HDM with
  FeynMaster 2}},  {\em JHEP} {\bf 06} (2021) 016,
  [\href{https://arxiv.org/pdf/2103.06281.pdf}{{\tt 2103.06281}}].

\bibitem{Fontes:2021znm}
D.~Fontes, M.~L\"{o}schner, J.~C. Rom\~{a}o, and J.~P. Silva, {\it {Leaks of CP
  violation in the real two-Higgs-doublet model}},  {\em Eur. Phys. J. C} {\bf
  81} (2021), no.~6 541, [\href{https://arxiv.org/pdf/2103.05002.pdf}{{\tt
  2103.05002}}].

\bibitem{Georgi:1978ri}
H.~Georgi and D.~V. Nanopoulos, {\it {Suppression of Flavor Changing Effects
  From Neutral Spinless Meson Exchange in Gauge Theories}},  {\em Phys. Lett.
  B} {\bf 82} (1979) 95--96.

\bibitem{Donoghue:1978cj}
J.~F. Donoghue and L.~F. Li, {\it {Properties of Charged Higgs Bosons}},  {\em
  Phys. Rev. D} {\bf 19} (1979) 945.

\bibitem{Botella:1994cs}
F.~J. Botella and J.~P. Silva, {\it {Jarlskog - like invariants for theories
  with scalars and fermions}},  {\em Phys. Rev. D} {\bf 51} (1995) 3870--3875,
  [\href{https://arxiv.org/pdf/9411288.pdf}{{\tt hep-ph/9411288}}].

\bibitem{Das:2019yad}
D.~Das and I.~Saha, {\it {Alignment limit in three Higgs-doublet models}},
  {\em Phys. Rev. D} {\bf 100} (2019), no.~3 035021,
  [\href{https://arxiv.org/pdf/1904.03970.pdf}{{\tt 1904.03970}}].

\bibitem{Boto:2021}
R.~Boto, {\it {Symmetry-constrained Multi-Higgs Doublet Models}},  Master's
  thesis, IST, Univ. Lisbon, 19 January 2021.

\bibitem{Grimus:2007if}
W.~Grimus, L.~Lavoura, O.~M. Ogreid, and P.~Osland, {\it {A Precision
  constraint on multi-Higgs-doublet models}},  {\em J. Phys.} {\bf G35} (2008)
  075001, [\href{https://arxiv.org/pdf/0711.4022.pdf}{{\tt 0711.4022}}].

\bibitem{Bento:2017eti}
M.~P. Bento, H.~E. Haber, J.~C. Rom\~{a}o, and J.~P. Silva, {\it {Multi-Higgs
  doublet models: physical parametrization, sum rules and unitarity bounds}},
  {\em JHEP} {\bf 11} (2017) 095,
  [\href{https://arxiv.org/pdf/1708.09408.pdf}{{\tt 1708.09408}}].

\bibitem{Bento:2022vsb}
M.~P. Bento, J.~C. Rom\~{a}o, and J.~P. Silva, {\it {Unitarity bounds for all
  symmetry-constrained 3HDMs}},  \href{https://arxiv.org/pdf/2204.13130.pdf}{{\tt
  2204.13130}}.

\bibitem{Ferreira:2010xe}
P.~M. Ferreira, L.~Lavoura, and J.~P. Silva, {\it {Renormalization-group
  constraints on Yukawa alignment in multi-Higgs-doublet models}},  {\em Phys.
  Lett. B} {\bf 688} (2010) 341--344,
  [\href{https://arxiv.org/abs/1001.2561}{{\tt 1001.2561}}].

\bibitem{Glashow:1976nt}
S.~L. Glashow and S.~Weinberg, {\it {Natural Conservation Laws for Neutral
  Currents}},  {\em Phys. Rev. D} {\bf 15} (1977) 1958.

\bibitem{Paschos:1976ay}
E.~A. Paschos, {\it {Diagonal Neutral Currents}},  {\em Phys. Rev. D} {\bf 15}
  (1977) 1966.


\bibitem{Yagyu:2016whx}
K.~Yagyu, {\it {Higgs boson couplings in multi-doublet models with natural
  flavour conservation}},  {\em Phys. Lett. B} {\bf 763} (2016) 102--107,
  [\href{https://arxiv.org/abs/1609.04590}{{\tt 1609.04590}}].





\bibitem{ATLAS-CONF-2018-031}
{\bf ATLAS Collaboration} Collaboration, A.~Collaboration, {\it {Combined
  measurements of Higgs boson production and decay using up to 80 fb$^{-1}$ of
  proton--proton collision data at $\sqrt{s}=$ 13 TeV collected with the ATLAS
  experiment}},  Tech. Rep. ATLAS-CONF-2018-031, CERN, Geneva, Jul, 2018.

\bibitem{Aranda:2019vda}
A.~Aranda, D.~Hern\'andez-Otero, J.~Hern\'andez-Sanchez, V.~Keus, S.~Moretti,
  D.~Rojas-Ciofalo, and T.~Shindou, {\it {Z$_3$ symmetric inert ( 2+1
  )-Higgs-doublet model}},  {\em Phys. Rev. D} {\bf 103} (2021), no.~1 015023,
  [\href{https://arxiv.org/pdf/1907.12470.pdf}{{\tt 1907.12470}}].

\bibitem{Chakraborti:2021bpy}
M.~Chakraborti, D.~Das, M.~Levy, S.~Mukherjee, and I.~Saha, {\it {Prospects of
  light charged scalars in a three Higgs doublet model with $Z_3$ symmetry}},
  {\em Phys. Rev. D} \textbf{104}, no.7, 075033 (2021)
  \href{https://arxiv.org/pdf/2104.08146.pdf}{{\tt 2104.08146}}.  

\bibitem{Baak:2014ora}
{\bf Gfitter Group} Collaboration, M.~Baak, J.~C\'uth, J.~Haller, A.~Hoecker,
  R.~Kogler, K.~M\"onig, M.~Schott, and J.~Stelzer, {\it {The global
  electroweak fit at NNLO and prospects for the LHC and ILC}},  {\em Eur. Phys.
  J. C} {\bf 74} (2014) 3046, [\href{https://arxiv.org/pdf/1407.3792.pdf}{{\tt
  1407.3792}}].

\bibitem{Akeroyd:2020nfj}
A.~G. Akeroyd, S.~Moretti, T.~Shindou, and M.~Song, {\it {CP asymmetries of
  ${\overline B}\to X_s/X_d\gamma$ in models with three Higgs doublets}},  {\em
  Phys. Rev. D} {\bf 103} (2021), no.~1 015035,
  [\href{https://arxiv.org/pdf/2009.05779.pdf}{{\tt 2009.05779}}].

\bibitem{Aad:2019mbh}
{\bf ATLAS} Collaboration, G.~Aad {\em et.~al.}, {\it {Combined measurements of
  Higgs boson production and decay using up to $80$ fb$^{-1}$ of proton-proton
  collision data at $\sqrt{s}=$ 13 TeV collected with the ATLAS experiment}},
  {\em Phys. Rev. D} {\bf 101} (2020), no.~1 012002,
  [\href{https://arxiv.org/pdf/1909.02845.pdf}{{\tt 1909.02845}}].

\bibitem{Spira:1995mt}
M.~Spira, {\it {HIGLU: A program for the calculation of the total Higgs
  production cross-section at hadron colliders via gluon fusion including QCD
  corrections}},  \href{https://arxiv.org/pdf/9510347.pdf}{{\tt
  hep-ph/9510347}}.

\bibitem{deFlorian:2016spz}
{\bf LHC Higgs Cross Section Working Group} Collaboration, D.~de~Florian {\em
  et.~al.}, {\it {Handbook of LHC Higgs Cross Sections: 4. Deciphering the
  Nature of the Higgs Sector}},  \href{https://arxiv.org/pdf/1610.07922.pdf}{{\tt
  1610.07922}}.

\bibitem{Bechtle:2020pkv}
P.~Bechtle, D.~Dercks, S.~Heinemeyer, T.~Klingl, T.~Stefaniak, G.~Weiglein, and
  J.~Wittbrodt, {\it {HiggsBounds-5: Testing Higgs Sectors in the LHC 13 TeV
  Era}},  {\em Eur. Phys. J. C} {\bf 80} (2020), no.~12 1211,
  [\href{https://arxiv.org/pdf/2006.06007.pdf}{{\tt 2006.06007}}].

\bibitem{Romao:1998sr}
J.~C. Romao and S.~Andringa, {\it Vector boson decays of the higgs boson},
  {\em Eur. Phys. J.} {\bf C7} (1999) 631--642,
  [\href{https://arxiv.org/pdf/9807536.pdf}{{\tt hep-ph/9807536}}].

\end{thebibliography}

\end{document}